\documentclass[MSNbibl,nameyear,dvips]{arxstspdf}
\usepackage{graphicx}
\usepackage{flushend}
\usepackage{stfloats}
\volume{28}
\issue{3}
\pubyear{2013}
\firstpage{376}
\lastpage{397}
\doi{10.1214/13-STS426} 

\makeatletter

\def\boldsymboll{\mathbf}
\renewcommand{\citep}[1]{(\citeauthor{#1}, \citeyear{#1})}

\setattribute{abstract}   {width}  {350pt}
\setattribute{keyword}    {width}  {350pt}
\makeatother

\begin{document}
\begin{frontmatter}

\title{Conflict Diagnostics in Directed Acyclic Graphs, with
Applications in Bayesian Evidence Synthesis}
\runtitle{Conflict Diagnostics in DAGs}
%
\begin{aug}

\author[a]{\fnms{Anne M.} \snm{Presanis}\corref{}\ead[label=e1]{anne.presanis@mrc-bsu.cam.ac.uk}},
\author[b]{\fnms{David} \snm{Ohlssen}},
\author[c]{\fnms{David J.} \snm{Spiegelhalter}}
\and
\author[d]{\fnms{Daniela} \snm{De Angelis}}

\runauthor{Presanis, Ohlssen, Spiegelhalter and De Angelis}

\affiliation{University Forvie Site, Novartis Pharmaceuticals Corporation, University of Cambridge Centre for Mathematical Sciences and
University Forvie Site}

\address[a]{Anne M. Presanis is Senior Investigator Statistician, MRC Biostatistics Unit,
Institute of Public Health,
University Forvie Site, Robinson Way, Cambridge CB2 0SR, United Kingdom
\printead{e1}.}
\address[b]{David Ohlssen is Senior Expert Methodologist, IIS Statistical Methodology,
Novartis Pharmaceuticals Corporation, One Health Plaza, Building 419, room 2157, East Hanover, New Jersey, 07936-1080, USA.}
\address[c]{David Spiegelhalter is Winton Professor for the Public Understanding of Risk,
Statistical Laboratory, University of Cambridge Centre for Mathematical Sciences, Wilberforce Road, Cambridge CB3 0WB, United Kingdom.}
\address[d]{Daniela De Angelis is Programme Leader, MRC Biostatistics Unit, Institute of Public Health,
University Forvie Site, Robinson Way, Cambridge CB2 0SR, United Kingdom.}
\end{aug}

%
\begin{abstract}
Complex stochastic models represented by directed acyclic\break graphs (DAGs)
are increasingly employed to synthesise multiple, imperfect and
disparate sources of evidence, to estimate quantities that are
difficult to measure directly. The various data sources are dependent
on shared parameters and hence have the potential to conflict with each
other, as well as with the model. In a Bayesian framework, the model
consists of three components: the prior distribution, the assumed form
of the likelihood and structural assumptions. Any of these components
may be incompatible with the observed data. The detection and
quantification of such conflict and of data sources that are
inconsistent with each other is therefore a crucial component of the
model criticism process. We first review Bayesian model criticism, with
a focus on conflict detection, before describing a general diagnostic
for detecting and quantifying conflict between the evidence in
different partitions of a DAG. The diagnostic is a $p$-value based on
splitting the information contributing to inference about a ``separator''
node or group of nodes into two independent groups and testing whether
the two groups result in the same inference about the separator
node(s). We illustrate the method with three comprehensive examples: an
evidence synthesis to estimate HIV prevalence; an evidence synthesis to
estimate influenza case-severity; and a hierarchical growth model for
rat weights.
\end{abstract}

%
\begin{keyword}
\kwd{Conflict}
\kwd{directed acyclic graph}
\kwd{evidence synthesis}
\kwd{graphical model}
\kwd{model criticism}\vspace*{-5pt}
\end{keyword}

\end{frontmatter}

\section{Introduction} \label{sec_intro}

Bayesian evidence synthesis methods combining multiple, imperfect and
disparate sources of data to estimate quantities that are challenging
to measure directly are becoming widespread (e.g., \cite{SpiegelhalterEtAl2004}; \cite{AdesSutton2006}). Although little data
may be available from which to directly estimate such quantities, there
may be plenty of \emph{indirect} information on related parameters that
can be expressed as functions of the key parameters of interest. The
synthesis of both direct and indirect data usually entails the
formulation of complex probabilistic models, where the dependency of
the data on the parameters is represented by a directed acyclic graph
(DAG). Recent examples can be found in the fields of ecology \citep
{ClarkEtAl2010}, biochemical kinetics (Henderson, Boys and Wilkinson, \citeyear{HendersonEtAl2010}),
environmental epidemiology \citep{JacksonEtAl2008},\break health technology
assessment \citep{WeltonEtAl2012}, mixed treatment comparisons \citep
{LuAdes2006} and infectious disease epidemiology \citep{BirrellEtAl2011}.

With modern software it has become reasonably straightforward to draw
inferences and make probabilistic predictions from such complex models.
However, with complexity also comes a vital requirement that the
conclusions of the model can be explained and justified, both for the
``owners'' of the model and any audience they wish to convince. There are
two main issues: First, to identify the essential \emph{drivers} of
inference, assessing sensitivity to data or assumptions. Second, to
judge whether the data are \emph{consistent} with each other or with
model assumptions. This assessment is crucial in syntheses of multiple
sources of evidence, where these sources are dependent on shared
parameters of interest and hence have the potential to conflict with
each other (\cite{LuAdes2006}; \cite{PresanisEtAl2008}). The evidence arising
from (i) prior distributions, (ii) the assumed form of the likelihood
and (iii) other structural/functional model assumptions also has the
potential to conflict with the different sources of data or with each
other. The existence of such inconsistency and/or sensitivity to model
assumptions would naturally lead to careful reexamination of the model
and data sources, and a further iteration of the inference and
model-criticism cycle recommended by \citet{Box1980}. \citet
{OHagan2003} reviews the connection between model checking and conflict
detection in the context of complex stochastic systems.

This paper focusses on the issue of detecting and measuring conflict,
particularly on diagnostics that are effective in the type of complex
DAG-based models that evidence synthesis requires for substantive
problems. Bayesian predictive $p$-values, in various guises, have been
widely employed as a natural measure to assess the consistency of the
components driving inference [e.g., \cite{Box1980}; \cite{GelmanEtAl1996}; \cite{BayarriCastellanos2007}]. \citet
{MarshallSpiegelhalter2007} proposed the idea of\vadjust{\goodbreak} ``node-splitting'' to
compare prior and likelihood in a hierarchical DAG-based model. A
general framework that unifies these various approaches has been
proposed (\cite{DahlEtAl2007}; G{\aa}se\-myr and Natvig (\citeyear{GasemyrNatvig2009})), but exploration of how
to apply these ideas in real-world complex problems remains limited
(\cite{DiasEtAl2010}; \cite{ScheelEtAl2011}). We review in Section~\ref{sec_back} the literature on Bayesian model checking, with a focus on
conflict detection. In Section~\ref{sec_methods} we describe a
generalisation of \citet{MarshallSpiegelhalter2007} to any node in a
DAG, with the aim of demonstrating how, in practice, such a diagnostic
may be usefully employed to detect and measure different types of
inconsistency in substantive problems. We give recommendations for
strategies to construct the diagnostic in different contexts and to
treat nuisance parameters. In Section~\ref{sec_egs} we then consider
three detailed examples: an evidence synthesis to estimate HIV
prevalence from many indirect data sources (\ref{sec_HIV}); an evidence
synthesis to estimate influenza severity (\ref{sec_flu}); and a
bivariate normal random effects model for rat weights, illustrating
multivariate conflict assessment (\ref{sec_rats}). We end with a
discussion in Section~\ref{sec_discuss}.

\section{Background} \label{sec_back}

\subsection{DAGs} \label{sec_DAGs}

It is now a standard procedure to use directed acyclic graphs to
represent the qualitative conditional independence assumptions of a
complex sto\-chastic model; see, for example, \citet{Lauritzen1996} for a
full description. The crucial idea is that each node in the graph
represents a stochastic quantity, related to other nodes through a
series of ``parent-child'' relationships to form a DAG. Using an
intuitive language in terms of familial relationships, the basic
assumption represented is that any node is conditionally independent of
its nondescendants given its parents. ``Founder'' nodes, that is, nodes
with no parents, are assigned a prior distribution. Given the directed
graphical structure, the joint distribution over all nodes is given by
the product of all the conditional distributions of each child given
its direct parents. Inference on DAGs is conducted when some nodes are
observed as data and the resulting joint and marginal posterior
distributions of the remaining nodes are needed. Substantial research
has led to efficient exact and simulation-based algorithms implemented
in various software (see \cite{CowellEtAl1999}, for a review), such as
Markov chain Monte Carlo [MCMC \cite{GamermanLopes2006}].\vadjust{\goodbreak}

%
\begin{figure*}

\includegraphics{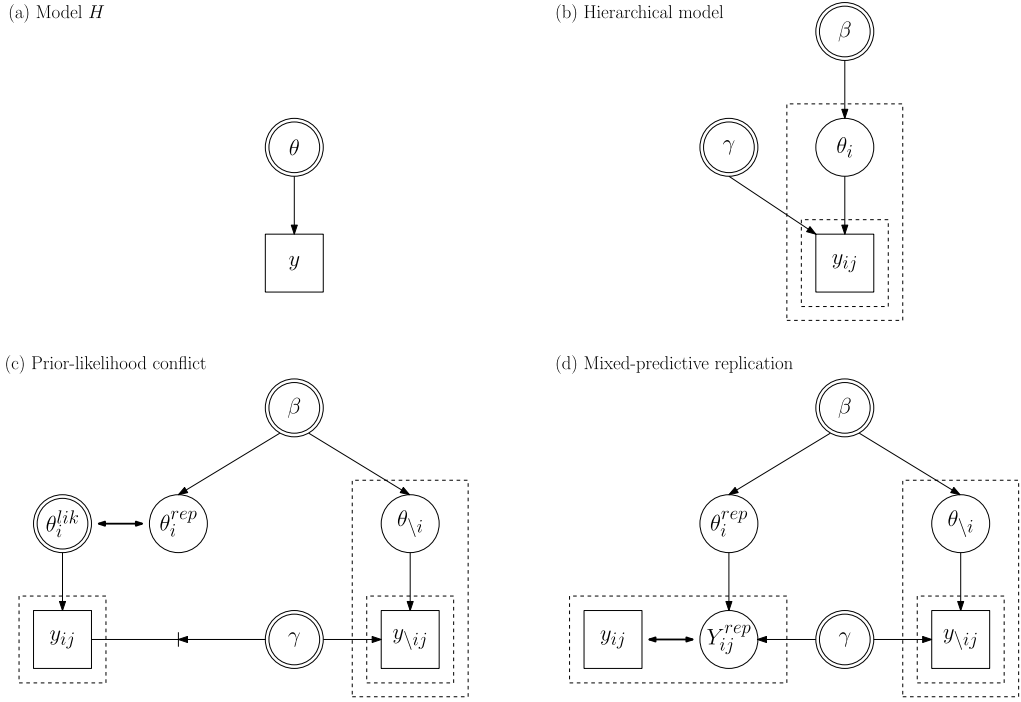}

\caption{Examples of DAGS showing \textup{(a)} a simple
model $H$; \textup{(b)} a hierarchical model; \textup{(c)} prior-likelihood comparison
via a node-split in model \textup{(b)}; and \textup{(d)} cross-validatory
mixed-predictive replication in model \textup{(b)}.}\label{fig_DAGevsyn}
\end{figure*}

Figure~\ref{fig_DAGevsyn}(a) shows a simple example of a DAG
representing a model $H$ for data $\boldsymboll{y}$ with parameters
$\bolds{\theta}$. The double circles represent the founder node
assigned a prior $p(\bolds{\theta} | H)$; the square represents
observations $\boldsymboll{y}$, a child node of $\bolds{\theta}$;
and the solid arrow represents a distributional assumption $\boldsymboll
{y} \sim p(\boldsymboll{y} | \bolds{\theta}, H)$. Figure~\ref{fig_DAGevsyn}(b) shows a slightly more complex hierarchical model,
where within units $i \in1\dvtx k$, the $n_i$ observations $y_{ij}, j \in
1\dvtx n_i$ are assumed drawn from a distribution with unit-specific
parameters $\theta_i$ and global parameters $\gamma$ (e.g., variances).
At the next level of hierarchy, the $\theta_i$ are drawn from
distributions with hyper-parameters $\beta$:
\begin{eqnarray*}
y_{ij} & \sim& p(y_{ij} | \theta_i, \gamma),\quad i
\in1\dvtx k, j \in 1\dvtx n_i,
\\
\theta_i & \sim& p(\theta_i | \beta),
\\
\beta, \gamma& \sim& p(\beta, \gamma).
\end{eqnarray*}
Repetition over and within units is represented by the dashed
rectangles, and the hyperparameters $\beta$ and $\gamma$ are the
founder nodes. Continuing the analogy of a family, $\gamma$ and $\theta
_i$ are co-parents of the data $\boldsymboll{y}_i$ and within groups
$i$, the data $\boldsymboll{y}_i$ are siblings.

\subsection{Node-Splitting}

\citet{MarshallSpiegelhalter2007} propose separating out the
contributions of prior and likelihood to a unit-level parameter $\theta
_i$ in a hierarchical model such as Figure~\ref{fig_DAGevsyn}(b), to
compare their consistency. They do so by splitting the DAG into two
independent partitions [Figure~\ref{fig_DAGevsyn}(c)]. The partition
representing the likelihood of the $i$th unit's data is formed by
drawing a replicate $\theta_i^{\mathrm{lik}} | \boldsymboll{y}_{i}$ from a
uniform reference prior, updated with the observations $\boldsymboll
{y}_{i}$, that is, from the posterior distribution generated by
$\boldsymboll{y}_{i}$ alone. This distribution is in effect the
``data-trans\-lated'' likelihood \citep{BoxTiao1973}. For the partition
representing the prior contribution to $\theta_i$, a replicate $\theta
_i^{\mathrm{rep}} | \boldsymboll{y}_{\setminus i}$ is drawn from the
``predictive-prior'' $p(\theta_i | \boldsymboll{y}_{\setminus i})$
where $\boldsymboll{y}_{\setminus i}$ denotes the remaining data aside
from unit $i$. The authors propose comparison of the posterior
distributions [represented by the double-headed arrow in Figure~\ref{fig_DAGevsyn}(c)] of the two replicates $\theta_i^{\mathrm{lik}}$ and $\theta
_i^{\mathrm{rep}}$ by defining a difference function, $\delta_i = \theta_i^{\mathrm{rep}}
- \theta_i^{\mathrm{lik}}$, then calculating the ``conflict $p$-value''
\[
c_{MS1,i} = \operatorname{Pr}(\delta_i \leq0 | \boldsymboll{y})
\]
in the case of a one-sided hypothesis test of departures towards
smaller $\theta_i$ than suggested by $\boldsymboll{y}_{i}$, or
\[
c_{MS2,i} = 2\min\bigl(\operatorname{Pr}(\delta_i\leq0 | \boldsymboll{y}),
1 - \operatorname{Pr}(\delta_i \leq0 | \boldsymboll{y})\bigr)
\]
if testing the null hypothesis $\delta_i = 0$ against $\delta_i \neq
0$. In the first case, values of $c_{MS1,i}$ close to either 0 or 1
indicate conflict. In the more general second case, a small value
represents a high level of conflict, so the ``conflict $p$-value'' is
possibly a misnomer: $c_{MS2,i}$ actually measures \emph{consistency}.
However, since the term has already been introduced in the literature,
we continue to refer throughout to the conflict $p$-value.

Finally, note also that this example is one where in order to
completely split the DAG into two independent partitions would require
splitting a vector of nodes, $\{\theta_i,\gamma\}$. However, if, for
example, $\gamma$ is a variance parameter in a normal hierarchical
model, we may not actually be directly interested in examining conflict
around $\gamma$, particularly if it is not strongly identifiable from a
unit $i$ alone. \citet{MarshallSpiegelhalter2007} therefore propose
treating such parameters as nuisance parameters, and ``cutting''
feedback from unit $i$ to $\gamma$ to prevent the data $\boldsymboll
{y}_{i}$ from influencing $\gamma$ [e.g., using the ``cut'' function in
the OpenBUGS software \citep{LunnEtAl2009}].\break A~cut in a DAG stops
information flow in one direction, as opposed to a node-split which
prevents information flow in both directions. This cut is represented
by the diode shape between $y_{ij}$ and $\gamma$ in Figure~\ref{fig_DAGevsyn}(c). The two replicates $\theta_i^{\mathrm{rep}}$ and $\theta
_i^{\mathrm{lik}}$ are therefore not entirely independent, since $\gamma$ may
still influence $\theta_i^{\mathrm{lik}}$; however, the authors believe any such
influence will be small.

\subsection{Bayesian Predictive Diagnostics} \label{sec_crit}

Bayesian predictive diagnostics to assess consistency are based on
comparison of a discrepancy statistic with a reference distribution.
The general setup is that of Figure~\ref{fig_DAGevsyn}(a), assuming a
model $H$ for $\boldsymboll{y}$ with parameters $\bolds{\theta}$:
\begin{eqnarray*}
\boldsymboll{y} & \sim& p(\boldsymboll{y} | \bolds{\theta}, H),
\\
\bolds{\theta} & \sim& p(\bolds{\theta} | H).
\end{eqnarray*}
To assess whether the observed data could have been generated by the
assumed model $H$, we compare the observed data $\boldsymboll{y}$, via a
test statistic $T(\boldsymboll{y})$, to a reference distribution $p_T\{
T(\boldsymboll{Y}^{\mathrm{rep}}) | H\}$ of the test statistic for hypothetical
(replicated) data $\boldsymboll{Y}^{\mathrm{rep}}$ under the assumed (null) model
$H$. The $p$-value defined by where the observed value of the test
statistic is located in the reference distribution measures the
compatibility of the model with the data. The reference distribution
depends on the way we assume the replicated data are generated from the
null model and therefore on the exact components of inference we wish
to compare. Various proposals have been made, broadly categorised into
prior-, posterior- and mixed-predictive approaches. Note that to be
able to interpret $p$-values in a meaningful way, the distribution of the
$p$-values under the null prior model $H$ is required.

The prior-predictive distribution \citep{Box1980}, in which the
parameters are integrated out with respect to the prior, is a natural
choice for the reference, as the $p$-values $\operatorname{Pr}\{T(\boldsymboll{Y}^{\mathrm{rep}})
\geq T(\boldsymboll{y}) | H\}$ are uniformly distributed under the null
prior model $H$. The approach assesses prior-data conflict, most
usefully in the case of informative priors. However, in the case of
improper priors, prior-predictive replication is not defined. In
practice, many analysts use very diffuse but proper priors to express
noninformativeness, in which case prior-data comparison is not
particularly useful, since almost any data will be plausible under such
priors. Other related approaches to assessing prior-data conflict
include: the adaptation by \citeauthor{EvansMoshonov2006}
(\citeyear{EvansMoshonov2006,EvansMoshonov2007})
of \citet{Box1980} to minimally sufficient statistics; the use of
logarithmic scoring rules (\cite
{Dawid1984}; Spiegelhalter et al. \citeyear{SpiegelhalterEtAl1993,SpiegelhalterEtAl1994}) to assess
conflict; and, more recently, the use of ratios of prior-to-posterior
distances under different priors, using Kullback-Leibler measures \citep
{Bousquet2008}.

The posterior-predictive distribution [e.g., \cite{Rubin1984}; \cite{GelmanEtAl1996}] was proposed as an alternative to
the prior-predictive distribution for use when improper priors are
employed. It results from integrating the parameters out with respect
to the posterior rather than prior distribution, thereby assessing
model-data rather than prior-data compatibility. Posterior-predictive
checks have become widespread \citep{GelmanEtAlBook2003}. However, the
$p$-values may not be uniformly distributed, since the data are used
twice, in obtaining both the posterior and the posterior-predictive
distribution (Bayarri and Berger \citeyear
{BayarriBerger1999,BayarriBerger2000}; \cite{RobinsEtAl2000}). Suggestions have
therefore been made to avoid the conservatism of posterior-predic\-tive
$p$-values, including alternative $p$-values that are closer to uniform,
but often difficult to compute, such as the conditional and partial
posterior-predic\-tive $p$-values (\citeauthor{BayarriBerger1999},
\citeyear{BayarriBerger1999,BayarriBerger2000};
Robins,\break van~der Vaart and Ventura (\citeyear{RobinsEtAl2000});
Bayarri and Cas\-tellanos (\citeyear{BayarriCastellanos2007})).
The predictive distributions in these approaches are defined by
integrating out the unknown parameters $\bolds{\theta}$ with
respect to posterior distributions that are, respectively, (i)
conditional on a sufficient statistic for $\bolds{\theta}$; and
(ii) constructed from the prior and from a likelihood defined to be
conditional on the observed value of a test statistic, $T(\boldsymboll
{y})$, so that the information in $T(\boldsymboll{y})$ contributing to
the posterior is ``removed'' before integrating out $\bolds{\theta
}$. Alternative ``double simulation'' approaches, post-processing
posterior-predictive $p$-values such\break that their distribution is uniform
(\cite{HjortEtAl2006}; \cite{Johnson2007}; \cite{SteinbakkStorvik2009}), require proper
priors and are computationally demanding.

The mixed-predictive distribution (\cite
{GelmanEtAl1996}; \cite{MarshallSpiegelhalter2007}), in the context of
hierarchical models such as Figure~\ref{fig_DAGevsyn}(b) with
hyperparameters $\beta$, integrates out the parameters $\bolds
{\theta} = \{\theta_i, i \in1\dvtx k\}$ with respect to what \citet
{MarshallSpiegelhalter2007} term the ``predic\-tive-prior'' distribution,
namely, $p^M(\bolds{\theta} | \boldsymboll{y}, H) =\break \int
p(\bolds{\theta} | \beta, H)\cdot p(\beta| \boldsymboll{y}, H) \,d\beta$.
This distribution is not the marginal posterior distribution of
$\bolds{\theta}$, but the distribution obtained by drawing
replicates $\bolds{\theta}^{\mathrm{rep}}$ from the marginal posterior
distribution of $\beta$. The parameters $\bolds{\theta}$ are then
integrated out, resulting in a mixed-predictive $p$-value that is still
conservative, but less so than the posterior-predictive $p$-value.

Posterior- and mixed-predictive approaches are often carried out in a
cross-validatory framework to avoid the double use of the data. Marshall and
Spie\-gelhalter (\citet{MarshallSpiegelhalter2007}) showed that under certain conditions, their
conflict $p$-value is equivalent to the cross-validatory mixed-predictive
$p$-value when this exists [Figure~\ref{fig_DAGevsyn}(c,d)]. In the
mixed-predictive approach, compatibility between the observations
$\boldsymboll{y}$ and the predictive distribution arising from both
likelihood $p(\boldsymboll{y} | \bolds{\theta})$ and prior
$p(\bolds{\theta} | \beta, H)$ is measured. In the conflict
approach, the prior is compared with the posterior arising from the
likelihood alone. Although the tests are mathematically the same under
the conditions described by \citet{MarshallSpiegelhalter2007}, the
mixed-predictive approach tests model/data compatibility, whereas the
conflict approach tests prior/likelihood compatibility. Other
cross-validatory approaches for model checking include the observed
relative surprise\break \citep{Evans1997} and a Bayesian influence
statistic\break
\citep{JacksonEtAl2012}.

\subsection{Node-Level Conflict Measures}

The predictive diagnostics of the previous section in general assess a
particular aspect of the whole model in comparison to the data. The
node-splitting idea of \citet{MarshallSpiegelhalter2007} is, by
contrast, a~method of assessing prior-likelihood conflict locally at a
particular node in the DAG. Other node-based diagnostics have been
proposed also, including that of \citet{OHagan2003}. He proposed
contrasting two sets of information informing a single node $\theta$,
where each set is summarised by a unimodal density or likelihood with
parameters $\bolds{\lambda}_a$ and $\bolds{\lambda}_b$
respectively. The author proposes first normalising both
densities/likelihoods to have unit maximum height, then considering the
height $z = p_a(x_z | \bolds{\lambda}_a) = p_b(x_z | \bolds
{\lambda}_b)$ at the point of intersection $x_z$ of the two curves
between the two modes. Taking $c_{\mathrm{OH}} =\break -2 \log(z)$ as the measure of
conflict, this will be high if the two densities/likelihoods have
little overlap. \citet{BayarriCastellanos2007} extend the partial
posterior predictive approach of Bayarri and Berger (\citeyear
{BayarriBerger1999,BayarriBerger2000}) to hierarchical models and, in
doing so, compare their method to several others, including \citeauthor{OHagan2003}'s
(\citeyear{OHagan2003}) conflict measure and Marshall and Spiegelhalter's (\citeyear{MarshallSpiegelhalter2007})
conflict $p$-value. They conclude that only their method and the conflict
$p$-value consistently detect conflict, noting that O'Hagan's measure may
be sensitive to the prior used and conservative, due to double use of the
data.

\citet{DahlEtAl2007} raise the same objection as \citet
{BayarriCastellanos2007} to the double use of data in O'Hagan's
measure, and therefore propose a variation on this measure, for the
simple normal hierarchical analysis of variance. They propose both a
data-splitting approach and a conflict measure at an internal node
$\theta$ of the DAG of the model, based on means and variances of
cross-validatory ``integrated posterior distributions'' (IPDs). These
IPDs are constructed by taking the information contribution to $\theta$
from one partition $\lambda_a$ of the DAG (either a prior or
likelihood), normalizing it to a density, and integrating it with
respect to the posterior distribution of the parameters $\lambda_b$ in
the other partition (analogous to posterior- or mixed-predictive
distributions for $\theta$, see next section for details). This is in
contrast to O'Hagan, who normalizes to unit height instead of to a
density. The authors derive the distributions of their conflict measure
for various data splittings under fixed known variances. They perform
an extensive simulation study for the case of unknown variances
assigned prior distributions, comparing their approach to O'Hagan's for
various data splittings and prior distributions.

\begin{figure}

\includegraphics[scale=0.99]{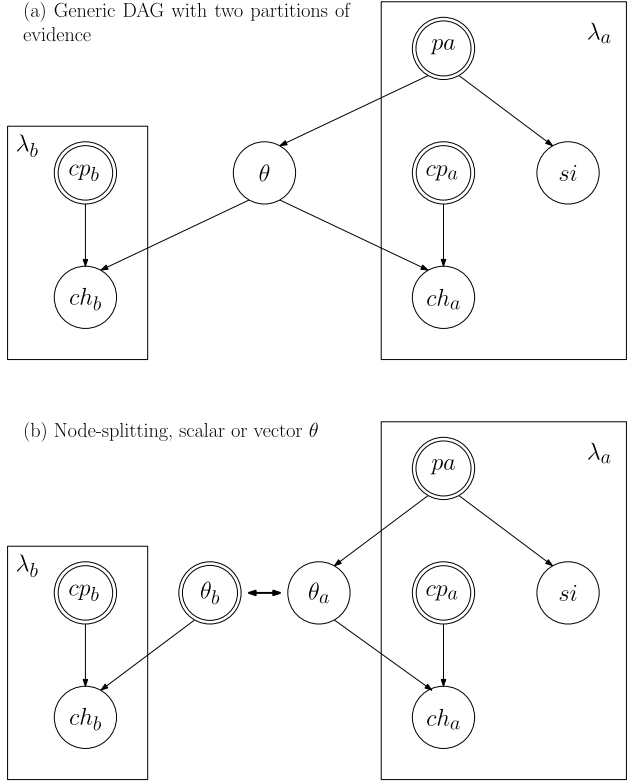}

\caption{DAG \textup{(a)} shows a generic model $H$,
with the evidence informing an internal ``separator'' node or set of
nodes $\bolds{\theta}$ partitioned into two groups, $\bolds
{\lambda}_a$ and $\bolds{\lambda}_b$, as in G{\aa}semyr and
  Natvig (\citeyear{GasemyrNatvig2009}). DAG \textup{(b)} shows the same generic model as \textup{(a)}, but
with the node(s) $\bolds{\theta}$ split into two copies, so that
the evidence that the two partitions $\bolds{\lambda}_a$ and
$\bolds{\lambda}_b$ provide about $\bolds{\theta}$ may be compared.}\label{fig_confDAGgeneric}
\end{figure}

\subsection{Unifying Approaches to Node-Level Conflict Diagnostics}

The models of Figure~\ref{fig_DAGevsyn} can be seen to be special cases
of the generic DAG in Figure~\ref{fig_confDAGgeneric}(a). The figure
shows a generic model $H$, with an internal ``separator'' node or set of
nodes $\bolds{\theta}$. The evidence informing $\bolds{\theta
}$ is partitioned into two groups, $\bolds{\lambda}_a$ and
$\bolds{\lambda}_b$. The set $\bolds{\lambda}_a$ contains
$\bolds{\theta}$'s parents $\boldsymboll{pa}$ and a subset of child
nodes $\boldsymboll{ch}_a$, with corresponding co-parents $\boldsymboll
{cp}_a$ and siblings $\boldsymboll{si}$. The partition $\bolds
{\lambda}_b$ contains the remaining child nodes $\boldsymboll{ch}_b$ and
corresponding co-parents $\boldsymboll{cp}_b$. The observed data
$\boldsymboll{y}$ are split into $\boldsymboll{y}_a$ and $\boldsymboll
{y}_b$, that are contained within the respective vectors of child
and/or sibling nodes $\{\boldsymboll{ch}_a, \boldsymboll{si}\}$ and
$\boldsymboll{ch}_b$. Each of the predictive diagnostics of Section~\ref{sec_crit}
are special cases of Figure~\ref{fig_confDAGgeneric}(a).
Figure~\ref{fig_confDAGgeneric}(b) shows the same generic model as (a),
but with the node(s) $\bolds{\theta}$ split into two copies, so
that the evidence that the two partitions $\bolds{\lambda}_a$ and
$\bolds{\lambda}_b$ provide about $\bolds{\theta}$ may be
compared. This figure represents a generalisation of \citeauthor{MarshallSpiegelhalter2007}'s
(\citeyear
{MarshallSpiegelhalter2007}) node-splitting\break approach to any internal
node(s) in a DAG.

The generic setup of Figure~\ref{fig_confDAGgeneric}(a) is described by
\citet{GasemyrNatvig2009}, who generalise their earlier work in \citet
{DahlEtAl2007}. They use the same cross-validatory IPDs as reference
distributions, but consider $p$-values based on tail areas of the
distributions, rather than a conflict measure based on means and
variances. The authors consider first conflict at a data node, in which
case the reference IPD is a posterior- or mixed-predictive distribution
of one partition of data conditional on the remaining data, that is, in
the cross-validatory setting of Figure~\ref{fig_confDAGgeneric}(a).
\citet{GasemyrNatvig2009} show that: for symmetric, unimodal IPDs,
their tail area conflict measure is equivalent to the measure based on
means and variances of \citet{DahlEtAl2007}; if the data are normally
distributed, their measure is equivalent to the cross-validatory
mixed-predictive $p$-value of \citet{MarshallSpiegelhalter2007}; and if
the IPDs are symmetric, unimodal and the data in the two partitions are
conditionally independent, their measure is equivalent to the partial
posterior predictive p-value of \citet{BayarriBerger2000}.

\citet{GasemyrNatvig2009} next consider conflict between two partitions
$\bolds{\lambda}_a$ and $\bolds{\lambda}_b$ of a DAG at an
internal node $\theta$ which is scalar. The IPDs are then predictive
distributions for $\theta$, conditional on the data in each partition:
\begin{eqnarray*}
p_a(\theta| \boldsymboll{y}_a) & = & \int f(\theta;
\bolds{\lambda} _a) p(\bolds{\lambda}_a |
\boldsymboll{y}_a) \,d\bolds{\lambda} _a,
\\
p_b(\theta| \boldsymboll{y}_b) & = & \int f(\theta;
\bolds{\lambda} _b) p(\bolds{\lambda}_b |
\boldsymboll{y}_b) \,d\bolds{\lambda}_b,
\end{eqnarray*}
where $f(\theta; \bolds{\lambda}_a)$ and $f(\theta; \bolds
{\lambda}_b)$ are densities proportional to the likelihood factors
informing $\theta$ in each of the two partitions, expressed as
functions of $\theta$. As in \citet{DahlEtAl2007}, \citet
{GasemyrNatvig2009} propose normalising the likelihood terms to
densities of $\theta$ conditional on the nodes in the partition,
$\bolds{\lambda}_a$ or $\bolds{\lambda}_b$. The authors take
a pair of independent samples $(\theta^*_a, \theta^*_b)$ from the two
predictive distributions $p_a(\theta| \boldsymboll{y}_a)$ and
$p_b(\theta| \boldsymboll{y}_b)$, and define $\delta= \theta^*_a -
\theta^*_b$. Their proposed conflict measures are tail area probabilities:
\begin{eqnarray*}
c_{GN3} & = & 1 - 2 \operatorname{min} \bigl\{\operatorname{Pr}(\delta\leq0), 1 - \operatorname{Pr}(\delta
\leq0) \bigr\},
\\
c_{GN4} & = & \operatorname{Pr} \bigl\{p_{\delta}(\delta| \boldsymboll{y}_a,
\boldsymboll {y}_b) \geq p_{\delta}(0 | \boldsymboll{y}_a,
\boldsymboll{y}_b) \bigr\},
\end{eqnarray*}
where $p_{\delta}$ is the posterior density of $\delta$. \citet
{GasemyrNatvig2009} demonstrate that the data-data conflict tail areas
they first considered are special cases of $c_{GN3}$ and $c_{GN4}$.
More generally, they also show that if the cumulative distribution
functions of $\theta$ (corresponding to the predictive densities $p_a$
and $p_b$, resp.) are normal, both $c_{GN3}$ and $c_{GN4}$ are
uniform pre-experimentally and are equivalent to each other. The
authors also extend their results to the general normal linear model
when the covariance matrix is fixed and known. Note that $c_{GN3}$
should be straightforward to obtain via simulation, using MCMC, for
example, whereas $c_{GN4}$ is much more computationally demanding,
requiring, for example, a kernel estimate of $p_{\delta}$. \citet
{GasemyrNatvig2009} note that $c_{GN3}$ is closely related to the
conflict $p$-value of \citet{MarshallSpiegelhalter2007}: by taking the
function $f(\theta; \bolds{\lambda}_b)$ to be proportional to the
likelihood of the data $\boldsymboll{y}_b$, \citet{GasemyrNatvig2009}
are implicitly assuming a uniform reference prior for $\theta$, whereas
\citet{MarshallSpiegelhalter2007} explicitly do so for the copy $\theta
_b$ [Figure~\ref{fig_confDAGgeneric}(b) versus (a)]. Finally, \citet
{GasemyrNatvig2009} extend their framework to multivariate node-splits
$\bolds{\theta}$, although their theoretical results are
restricted to cases where the two predictive distributions are
multivariate normal, and to general normal linear models with known covariances.

In a slightly different approach, that complements and is related to
the conflict measures summarised here, \citet{ScheelEtAl2011} propose a
diagnostic plot to visualise conflict at any particular node $\theta$
in a DAG. The authors define a ``local prior'' for $\theta$ conditional
on its parents and a ``lifted likelihood'' coming from $\theta$'s
children, conditional on both $\theta$ and the co-parents of $\theta$'s
children. The ``local critique plot'' then examines where the marginal
posterior of $\theta$ lies, relative to both the local prior and the
lifted likelihood.

\section{Extending the Conflict $p$-Value to Any Node} \label{sec_methods}

While the conflict measures summarised in the previous section are
useful tools, the general framework introduced by \citet{DahlEtAl2007}
and \citet{GasemyrNatvig2009} uses idealised examples, such as normal
models or general normal linear models with fixed covariances, to
demonstrate uniformity of $p$-values. Furthermore, many of the other
measures, such as post-processed $p$-values, are computationally
expensive. In the context of complex evidence syntheses, a diagnostic
for conflict is required that is both easy to compute and applicable in
the more complex probabilistic models typical of evidence synthesis. We
emphasise that in presenting a generalisation of the conflict $p$-value
of \citet{MarshallSpiegelhalter2007} to any ``separator'' node(s)
$\bolds{\theta}$ in a DAG [Figure~\ref{fig_confDAGgeneric}(b)], we
are not aiming to prove uniformity of $p$-values in specific cases.
Instead, we aim to present a framework for conflict detection in \emph
{practice}, demonstrating the utility of such methods in substantive,
realistic examples.

In the context of Figure~\ref{fig_confDAGgeneric}(b), the evidence
informing $\bolds{\theta}$ is comprised of the information coming
from each of $\bolds{\theta}$'s neighbours in the DAG. This
information is generally in the form of either: (i) a (potentially
predictive) prior distribution from $\bolds{\theta}$'s parents and
siblings, that may include likelihood terms from $\boldsymboll{si}$; or
(ii) likelihood contributions from $\bolds{\theta}$'s children
combined with priors for the co-parents. Note that Figure~\ref{fig_confDAGgeneric}(b), although general, could be generalised even
further if, for example, the co-parents have other children that are
not children of $\bolds{\theta}$, but also contain likelihood
terms, indirectly informing~$\bolds{\theta}$.

\begin{figure}

\includegraphics{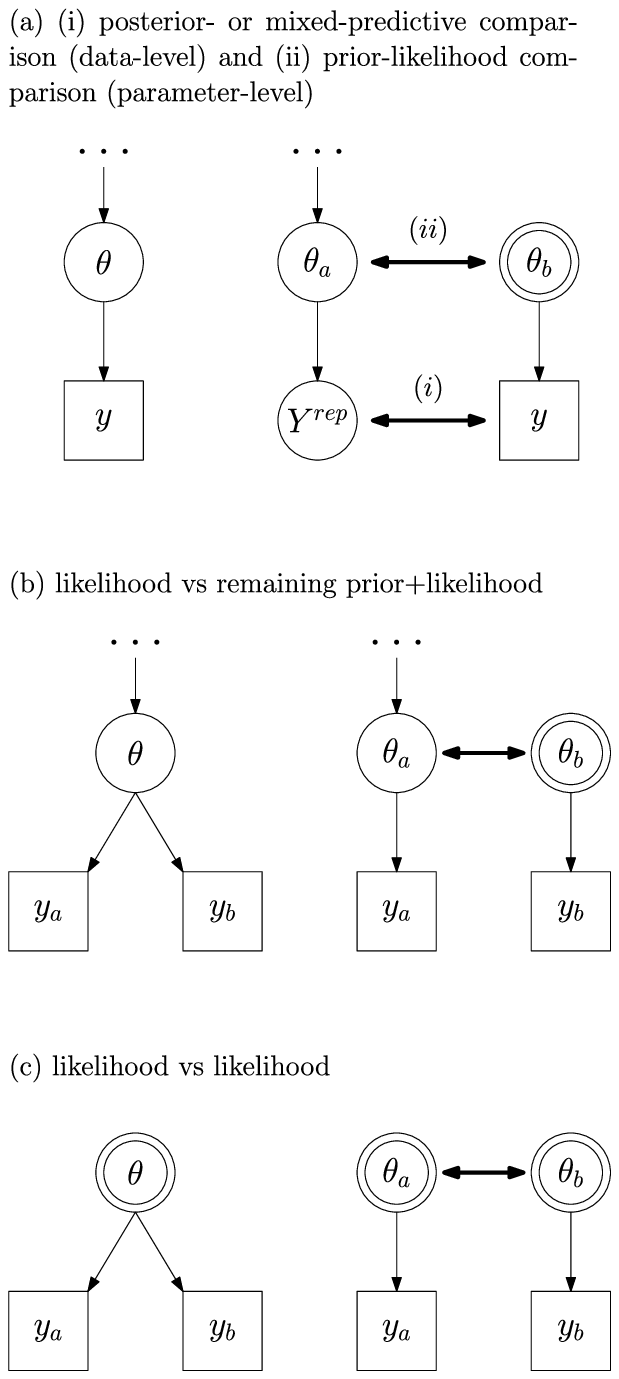}

\caption{Examples of node-splits contrasting \textup{(a)}
\textup{(i)} data and predictive distribution or \textup{(ii)} prior and likelihood; \textup{(b)}
a likelihood term and the remaining combined prior and likelihood; and
\textup{(c)} two likelihood terms, assuming a reference prior for each $\theta$.
In each example, the original model is on the left and the node-split
is shown on the right.}\label{fig_simpleEGs}
\end{figure}

The evidence surrounding $\bolds{\theta}$ is split into two
independent groups, and we compare the inferences about $\bolds
{\theta}$ resulting from each of the two partitions by comparing the
two (independent) posterior distributions $p(\bolds{\theta}_a |
\boldsymboll{y}_a)$ and $p(\bolds{\theta}_b | \boldsymboll{y}_b)$.
We assess our null hypothesis that $\bolds{\theta}_a = \bolds
{\theta}_b$. Our measure of conflict can be designed to reflect
differences between data and a model, between different data sources,
or between a prior and a likelihood. Simple examples of different types
of node-splits are given in Figure~\ref{fig_simpleEGs}, each of which
is a special case of Figure~\ref{fig_confDAGgeneric}(b). The examples
include comparison between a likelihood and the remaining combined
prior and likelihood, appropriate when questioning part of the data
contributing to a node [Figure~\ref{fig_simpleEGs}(b)] and a likelihood
vs likelihood comparison [Figure~\ref{fig_simpleEGs}(c)], directly
contrasting sources of evidence without taking into account a prior.
%

\subsection{Choice of a Reference Prior} \label{sec_refprior}
The question of how to create the two posterior distributions to
compare depends on the precise comparison we wish to make, and hence
the specific node-splits we construct. In particular, the creation of
posterior distributions depends partly on the choice of a reference
prior to use for $\bolds{\theta}_b$ or, indeed, for both
$\bolds{\theta}_a$ and $\bolds{\theta}_b$ in the case of
node-splits of a form such as Figure~\ref{fig_simpleEGs}(c). The aim is
to choose a prior that turns what is effectively a likelihood term
[$p(\boldsymboll{y}_b | \boldsymboll{cp}_b, \bolds{\theta}_b)$,
where $\boldsymboll{cp}_b$ is a vector of $\bolds{\theta}_b$'s
co-parents] into a posterior distribution, without the prior itself
influencing the posterior. We therefore follow \citet{BoxTiao1973} and
\citet{Kass1990} in adopting uniform priors for a transformation
$h(\bolds{\theta}_b)$ of $\bolds{\theta}_b$ to a scale such
that the uniform prior is appropriate. The posterior distribution is
then effectively the ``data-translated likelihood'' \citep
{BoxTiao1973}. As noted previously, \citet{MarshallSpiegelhalter2007}
showed that under certain conditions, choosing a uniform prior results
in the conflict $p$-value and the cross-validatory mixed-predictive
$p$-value coinciding when the latter exists. They note also that in other
situations, \citet{BoxTiao1973} showed that the Jeffreys' priors widely
used as ``noninformative'' priors are equivalent to a uniform prior on an
appropriate transformation of $\bolds{\theta}_b$. We therefore
recommend use of a Jeffreys' prior in general, although note that for
some node-splits, choice of a ``noninformative'' prior may not be so
straightforward. For some comparisons, we may not be able to assign a
Jeffreys' prior or there may not be a natural choice of reference
prior. We then rely on the likelihood in a specific partition
dominating any prior chosen for founder nodes in that partition, though
this assumption should be assessed, for example, through sensitivity
analyses to the prior (see Section~\ref{sec_discuss} for further
discussion of this).

\subsection{Comparison of Two Posterior Distributions} \label{sec_compare}

Considering for now scalar $\theta$, to test the point-null hypothesis
that $\theta_a = \theta_b$, Bayes factors could be employed. However,
the approach is known to be difficult because of the high dependence on
the precise form of the reference priors, as well as hard to evaluate
using MCMC. Instead, we prefer to consider the plausibility of the
hypothesis either directly if $\theta$ takes discrete values or using a
$p$-value if the support of $\theta$ is continuous.

\subsubsection{Conflict at a discrete node}

If $\theta$ takes values in $0,1,\ldots,K$, then we can directly
evaluate
\begin{eqnarray*}
c &=& p(\theta_a = \theta_b | \boldsymboll{y}_a,
\boldsymboll{y}_b)\\
& =& \sum_{k=0}^K
p(\theta_a = k | \boldsymboll{y}_a) p(
\theta_b = k | \boldsymboll{y}_b).
\end{eqnarray*}
As a simple example, consider a disease with known prevalence $\pi$ and
denote the event of having the disease by $\theta= 1$ and of no
disease by $\theta= 0$. The prior probability of having the disease is
therefore $p(\theta= 1) = \pi$. A diagnostic test $Y$ has sensitivity
$s$ and specificity $t$, so that $p(Y = 1 | \theta= 1) = s$ and $p(Y =
0 | \theta= 0) = t$. If a positive test result $y = 1$ is observed,
then the posterior probability of the disease is $p(\theta= 1 | y = 1)
= \pi s /  \{\pi s + (1-\pi)(1-t) \}$.

If we wish to assess the conflict between the prior $\theta\sim\operatorname
{Bernoulli}(\pi)$ and the likelihood of observing a positive test
result given $\theta$, $p(y = 1 | \theta)$, then, in this case, $K =
1$, $\theta_a$ is the copy representing the prior, $\theta_b$ is the
copy representing the likelihood, $\boldsymboll{y}_a$ is the empty set
and $\boldsymboll{y}_b = y$ [see also Figure~\ref{fig_simpleEGs}(a)(ii)]. We assume a reference prior $p(\theta_b = 1) =
0.5$, and so obtain a reference posterior $p(\theta_b = 1 | y = 1) = s
/  \{s + (1-t) \}$. The conflict measure is then
\[
c = \bigl\{\pi s + (1-\pi) (1-t) \bigr\} / \bigl\{s + (1-t) \bigr\}.
\]
For example, if a diagnostic test has sensitivity and specificity 0.9,
then the conflict when observing a positive test result is $c = 0.1 +
0.8\pi$, which has a minimum of 0.1 for very rare diseases.

\subsubsection{Conflict at a continuous node}

For $\theta$ with continuous support, we can no longer calculate\break
$p(\theta_a = \theta_b | \boldsymboll{y}_a, \boldsymboll{y}_b)$ directly.
Instead, we consider the posterior probability of $\delta= h(\theta_a)
- h(\theta_b)$, where $h$ is the function (Section~\ref{sec_refprior})
we choose that turns $\theta$ into a parameter for which it is
reasonable to assume a uniform prior. To judge whether the null
hypothesis $\delta= 0$ is plausible, and as do \citet
{GasemyrNatvig2009}, we adapt \citeauthor{Box1980}'s (\citeyear{Box1980}) suggestion of
calculating the probability that a predictive density is smaller than
the predictive density at the observed data, to calculating the
probability that the \emph{posterior} density of $\delta$ is smaller
than that at 0, {that is},
\[
c = \operatorname{Pr} \bigl\{ p_{\delta}(\delta| \boldsymboll{y}_a,
\boldsymboll{y}_b) < p_{\delta
}(0 | \boldsymboll{y}_a,
\boldsymboll{y}_b) \bigr\}.
\]
This can also be interpreted as adopting a log scoring rule
(\citeauthor{SpiegelhalterEtAl1993},
\citeyear
{SpiegelhalterEtAl1993,SpiegelhalterEtAl1994}; \cite{GneitingRaftery2007}) for
$\delta$:
\[
c = \operatorname{Pr} \bigl\{ - \log p_{\delta}(\delta| \boldsymboll{y}_a,
\boldsymboll{y}_b) > - \log p_{\delta}(0 |
\boldsymboll{y}_a, \boldsymboll{y}_b) \bigr\},
\]
which we can think of as the predictive probability of getting a higher
penalty than if we believe the null hypothesis that $\delta= 0$.

A problem is the need for evaluation of the posterior density $p_{\delta
}(\delta| \boldsymboll{y}_a, \boldsymboll{y}_b)$, which is not available analytically
in any but the simplest of examples. However, the Jeffreys'
transformation $h(\cdot)$ to a location parameter on the real line may
also ensure that the posterior distribution of $\delta$ is symmetric
and unimodal. In this case, the conflict measure is simply double the
tail-area beyond~$0$:
\[
c = 2 \times\min \bigl\{ \operatorname{Pr}(\delta> 0 | \boldsymboll{y}_a,
\boldsymboll{y}_b), 1 - \operatorname{Pr}(\delta> 0 | \boldsymboll{y}_a,
\boldsymboll{y}_b) \bigr\},
\]
which is easily implemented using MCMC by counting the number of times
$\theta_a$ exceeds $\theta_b$ in the sample. As \citet
{GasemyrNatvig2009} have shown, if $h(\theta_a), h(\theta_b)$---and
hence $\delta$ if $\theta_a$ and $\theta_b$ are indepen\-dent---are
normally distributed, then the $p$-value is uniformly distributed {a
priori}. A value of $c$ close to $0$ therefore indicates a low degree
of consistency between the posterior distributions of $\theta_a$ and
$\theta_b$.

If the distribution is not symmetric, a one-sided tail area may be
more\vadjust{\goodbreak}
appropriate. As we will see in some of the examples of Section~\ref{sec_egs},
we may in any case be interested in one-tailed tests $\delta
\geq0$ or $\delta\leq0$, in which case asymmetry is not a problem.
Clearly, if the distribution is multi-modal, the tail area is not an
appropriate measure of where $0$ lies in the posterior distribution of
$\delta$ \citep{EvansJang2010}. Then \citep{GasemyrNatvig2009} we may
consider using a kernel density estimate of MCMC samples from the
posterior distribution to empirically obtain $c = \operatorname{Pr}  \{ p_{\delta
}(\delta| \boldsymboll{y}_a, \boldsymboll{y}_b) < p_{\delta}(0 | \boldsymboll{y}_a,
\boldsymboll{y}_b)  \}$, though this will clearly be dependent on the choice of
bandwidth and kernel. We defer further discussion of these issues to
Section~\ref{sec_discuss}.

\subsubsection{Conflict at multiple continuous nodes} \label{sec_multi}

As seen from the example in \citet{MarshallSpiegelhalter2007} and
Figure~\ref{fig_DAGevsyn}(c) of this paper, to obtain completely
independent partitions of evidence in a DAG, often (particularly in
hierarchical models) a vector of nodes would need to be split. How the
DAG should be split will be context-dependent: either we are interested
in comparing what we can infer about all the separator nodes in the
vector from the two partitions of evidence or some of the separator
nodes are nuisance parameters [such as the variance $\gamma$ in Figure~\ref{fig_DAGevsyn}(c)] in which we are not directly interested. If the
latter is the case, we can impose a cut such as the one shown in Figure~\ref{fig_DAGevsyn}(c) to prevent information from one partition
influencing the nuisance parameters. In the former case, we can split
each node in the vector and examine the posterior distribution of
$\bolds{\delta} = \boldsymboll{h}_a - \boldsymboll{h}_b =  \{
h_1(\theta_{a1}),\ldots,h_k(\theta_{ak}) \}^T -  \{h_1(\theta
_{b1}),\ldots,h_k(\theta_{bk}) \}^T$, where $k$ is the length of
the vector and the functions $h_1,\ldots,h_k$ are the appropriate
Jeffreys' transformations. The key question is then how to calculate a
multivariate $p$-value to test $\bolds{\delta} = \boldsymboll{0}$. We
consider three options, dependent on the posterior distribution of
$\bolds{\delta}$:

\renewcommand{\labelenumi}{\roman{enumi}.}
\begin{longlist}[(iii)]
\item[(i)] If we are willing to assume multivariate normality for the
posterior $p_{\bolds{\delta}}(\bolds{\delta} | \boldsymboll
{y}_a, \boldsymboll{y}_b)$, and denoting the posterior expectation and
covariance of $\bolds{\delta}$ by $\mathbb{E}_p$ and $\operatorname
{Cov}_p$, respectively, then \citep{GasemyrNatvig2009} the standardised
discrepancy measure
\[
\Delta= \mathbb{E}_p(\bolds{\delta})^T
\operatorname{Cov}_p(\bolds {\delta})^{-1}\mathbb{E}_p(
\bolds{\delta})
\]
may be compared with a $\mathcal{X}^2$ distribution with $k$ degrees of
freedom to obtain a conflict measure $c = 1 - \operatorname{Pr} \{\mathcal{X}^2_k
\leq\Delta \}$.

\item[(ii)] If we are not willing to assume multivariate normality, but the
posterior density $p_{\bolds{\delta}}(\bolds{\delta} |
\boldsymboll{y}_a, \boldsymboll{y}_b)$ is still symmetric and uni-modal,
we can sample points from the posterior (e.g., using MCMC) and for each
sample~$\bolds{\delta}_i$, calculate its Mahalanobis distance from
their mean
\[
\Delta_i = \bigl\{\bolds{\delta}_i -
\mathbb{E}_p(\bolds{\delta})\bigr\} ^T
\operatorname{Cov}_p(\bolds{\delta})^{-1}\bigl\{\bolds{
\delta}_i - \mathbb{E}_p(\bolds{\delta})\bigr\}.
\]
Then a conflict measure is $c = \operatorname{Pr} \{\Delta_i > \Delta \}$,
the proportion over the MCMC sample of points that are further away
from the mean than is $\boldsymboll{0}$. This is a means of calculating
the multivariate tail area probability, analogous to counting the
number of times in the MCMC sample $\delta$ is greater than $0$ in the
univariate case.

\item[(iii)] Finally, if the posterior distribution is skew and/or
multi-modal, we could, as in the univariate case, obtain a kernel
density estimate of the posterior based on the MCMC samples and use the
estimate to calculate the probability that the posterior density at
$\bolds{\delta}$ is less than (at a lower contour than) at
$\boldsymboll{0}$:
\[
c = \operatorname{Pr} \bigl\{ p_{\bolds{\delta}}(\bolds{\delta} | \boldsymboll
{y}_a, \boldsymboll{y}_b) < p_{\bolds{\delta}}(
\boldsymboll{0} | \boldsymboll{y}_a, \boldsymboll{y}_b)
\bigr\},
\]
that is, that $\boldsymboll{0}$ lies in the tail of the distribution.
\end{longlist}
\renewcommand{\labelenumi}{\arabic{enumi}.}
Note that the third approach will again be dependent on the choice of
bandwidth and kernel.

\section{Examples} \label{sec_egs}

We now illustrate the use of conflict $p$-values to detect conflict in a
series of three increasingly complex evidence syntheses. All analyses
were carried out in OpenBUGS 3.2.2 \citep{LunnEtAl2009} and R 2.15.0
\mbox{\citep{Rproject2005}}.

\begin{figure}

\includegraphics{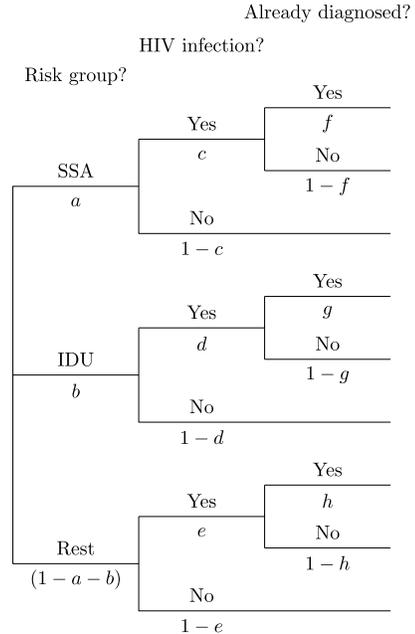}

\caption{The HIV example: probability tree showing
the epidemiological model from Ades and Cliffe (\citeyear{AdesCliffe2002}). ``SSA'' denotes
women born in sub-Saharan Africa and ``IDU'' denotes injecting drug using women.}\label{fig_HIVtree}
\end{figure}

\subsection{HIV Example} \label{sec_HIV}

\begin{figure}

\includegraphics{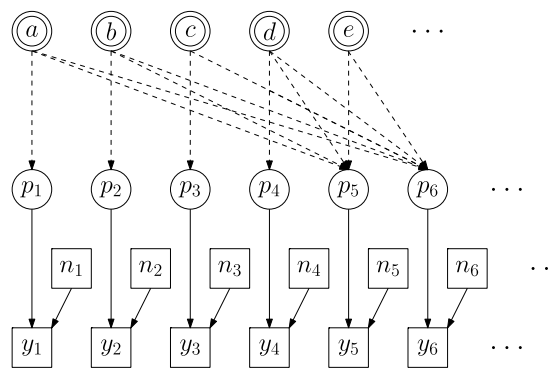}

\caption{The HIV example: DAG showing part of the
model. Note the functional relationships represented by the dashed arrows.}\label{fig_HIVdag}
\end{figure}

\citet{AdesCliffe2002} proposed a Bayesian synthesis of multiple
sources of evidence to examine alternative strategies for screening HIV
in prenatal clinics, with the specific aim of deciding whether to
universally screen or to use targeted screening of high risk groups.
Figure~\ref{fig_HIVtree} shows the epidemiological part of their model,
with the ``basic parameters'' $a$ to $h$ to be estimated. These letters
represent the probabilities of women in prenatal clinics being in a
particular risk group [$a$, $b$ and $1-a-b$, representing,
respectively, women born in sub-Saharan Africa (SSA), women injecting
drugs (IDU) and the remaining women]; being HIV infected ($c,d$ and
$e$, respectively, for the three risk groups); and being already diagnosed
prior to clinic visit ($f,g$ and $h$, resp., for HIV positive
women in each of the three risk groups).
Direct evidence is only available for a limited number of these
parameters, but there is also indirect evidence informing functions
(``functional parameters'') of $a$ to $h$ as well as of an extra basic
parameter $w$. Direct evidence is defined as a study with the aim of
measuring a basic parameter. Indirect evidence is provided by the other
studies through the logical functions that link the basic parameters.
Table~\ref{tab_HIVdata} shows the available data and the parameters,
both basic and functional, that these inform, while Figure~\ref{fig_HIVdag} shows a DAG of part of the model, demonstrating both the
distributional and functional relationships. The basic parameters are
the founder nodes to which prior distributions are assigned, whereas
the functional parameters $p_1, \ldots, p_{12}$ are the probabilities
informed directly by each of the 12 data sources.

\begin{table*}
\caption{HIV example: data sources and the
parameters they inform. ``SSA'' denotes sub-Saharan Africa and ``IDU''
denotes injecting drug using. ``Seroprevalence'' is the prevalence of HIV
antibodies in blood samples from a ``sero-survey''}\label{tab_HIVdata}
\begin{tabular*}{\textwidth}{@{\extracolsep{\fill}}lp{6.7cm}cccc@{}}
\hline
 &  &  & \textbf{Data} & & \\
\textbf{Source}&\multicolumn{1}{c}{\textbf{Description of data}} &\textbf{Parameter} & \multicolumn{1}{c}{$\bolds{y}$} & \multicolumn{1}{c}{$\bolds{n}$} &  \multicolumn{1}{c}{$\bolds{y / n}$} \\
\hline
\phantom{0}1 & Proportion of women born in SSA, 1999 &
$p_1 = a$
& $11\mbox{,}044$ & 104\mbox{,}577 & 0.106 \\
\phantom{0}2 & Proportion of women who are IDU in the last 5 years &
$p_2 = b$
& 12 & 882 & 0.014 \\
\phantom{0}3 & HIV prevalence in women born in SSA, 1997--1998 &
$p_3 = c$
& 252 & 15,428 & 0.016 \\
\phantom{0}4 & HIV prevalence in IDU women, 1997--1999 &
$p_4 = d$
& 10 & 473 & 0.021 \\
\phantom{0}5 & HIV prevalence in women not born in SSA, 1997--1998 &
$p_5 = \frac{db + e(1-a-b)}{1-a}$
& 74 & 136,139 & 0.001 \\
\phantom{0}6 & HIV seroprevalence in pregnant women, 1999 &
$p_6 = ca + db + e(1-a-b)$
& 254 & 102,287 & 0.002 \\
\phantom{0}7 & Diagnosed HIV in SSA-born women as a proportion of all diagnosed
HIV, 1999 &
$p_7 = \frac{fca}{fca + gdb + he(1-a-b)}$
& 43 & 60 & 0.717 \\
\phantom{0}8 & Diagnosed HIV in IDU women as a proportion of diagnosed HIV in
non-SSA-born women, 1999 &
$p_8 = \frac{gdb}{gdb + he(1-a-b)}$
& 4 & 17 & 0.235 \\
\phantom{0}9 & Overall proportion of HIV diagnosed &
$p_9 = \frac{fca+gdb+he(1-a-b)}{ca + db + e(1-a-b)}$
& 87 & 254 & 0.343 \\
10 & Proportion of infected IDU women diagnosed, 1999 &
$p_{10} = g$
& 12 & 15 & 0.800 \\
11 & Proportion of infected SSA-born women with serotype B, 1997--1998 &
$p_{11} = w$
& 14 & 118 & 0.119 \\
12 & Proportion of infected non-SSA-born women with serotype B,
1997--1998, assuming that $100\%$ of infected IDU women have serotype B
and that infected non-SSA-born non-IDU women have the same prevalence
of serotype B as infected SSA-born women &
$p_{12} = \frac{db + we(1-a-b)}{db + e(1-a-b)}$
& 5 & 31 & 0.161 \\
\hline
\end{tabular*}
\end{table*}

\begin{figure}

\includegraphics{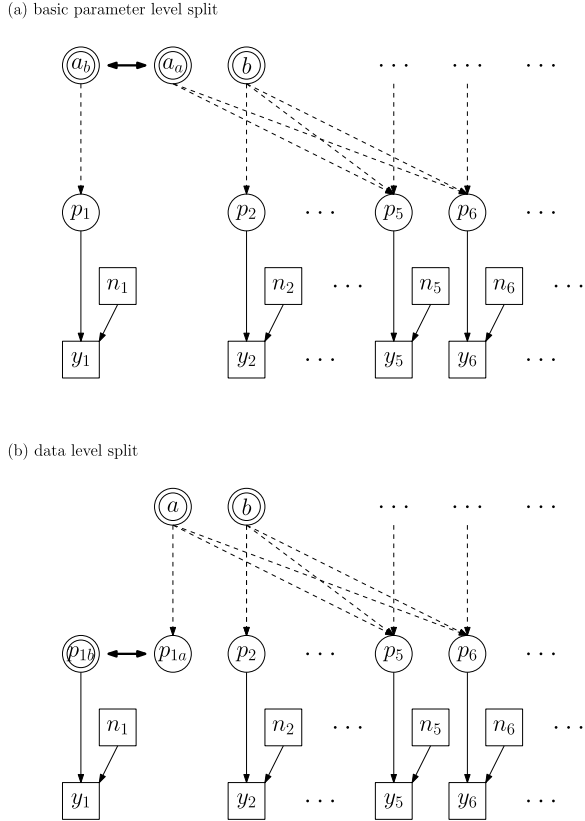}

\caption{The HIV example: DAG showing the two
types of node splits, one example of each. In \textup{(a)}, the node $a$ is
split to reflect direct ($a_b$) versus indirect ($a_a$) evidence. In
\textup{(b)}, the node $p_1$ is split to reflect prior (all the indirect
evidence on $p_{1a}$) versus likelihood (direct evidence on $p_{1b}$)
in a cross-validation approach.}\label{fig_HIVsplits}
\end{figure}

The effect of the direct and indirect evidence on inference may be
compared using two different types of node-splits: one at the level of
the basic parameters and the second at the level of the probabilities
$p_i, i \in1, \ldots, 12$. The two types of node-splits are shown in
Figure~\ref{fig_HIVsplits}. The first node-split is carried out for
each of the six basic parameters $\theta\in\{a,b,c,d,g,w\}$ for which
direct data are available (studies 1 to 4, 10 and 11). The indirect
evidence comprises all the remaining studies and the functional
relationships assumed. The second type of node-split compares, for each
$i \in1, \ldots, 12$, the direct evidence on $p_i$ provided by study
$i$ with the indirect evidence provided by the remaining studies, {that is}, a data-level cross-validation.

We adopt Jeffreys' prior for the nodes representing direct evidence in
each case:
$\theta_b \sim\operatorname{Beta}({1}/{2},{1}/{2})$ for each
$\theta\in\{a,b,c,d,g,w\}$ in the first set of node-splits; and
$p_{ib} \sim\operatorname{Beta}({1}/{2},{1}/{2})$ in
partition $b$ in the second set of node-splits [Figure~\ref{fig_HIVsplits}(b)]. In both node-splits, the conflict $p$-values are
two-tailed since we are testing for nonequality. Since the posterior
densities of each difference function $\delta$ appear symmetric and
uni-modal, the $p$-values are defined by taking twice the proportion of
MCMC samples where $\delta\geq0$ or $\delta< 0$, whichever is
smaller. The results are based on two chains of $20\mbox{,}000$ MCMC
iterations each, after discarding a burn-in of $20\mbox{,}000$ iterations.

%

\begin{figure*}

\includegraphics{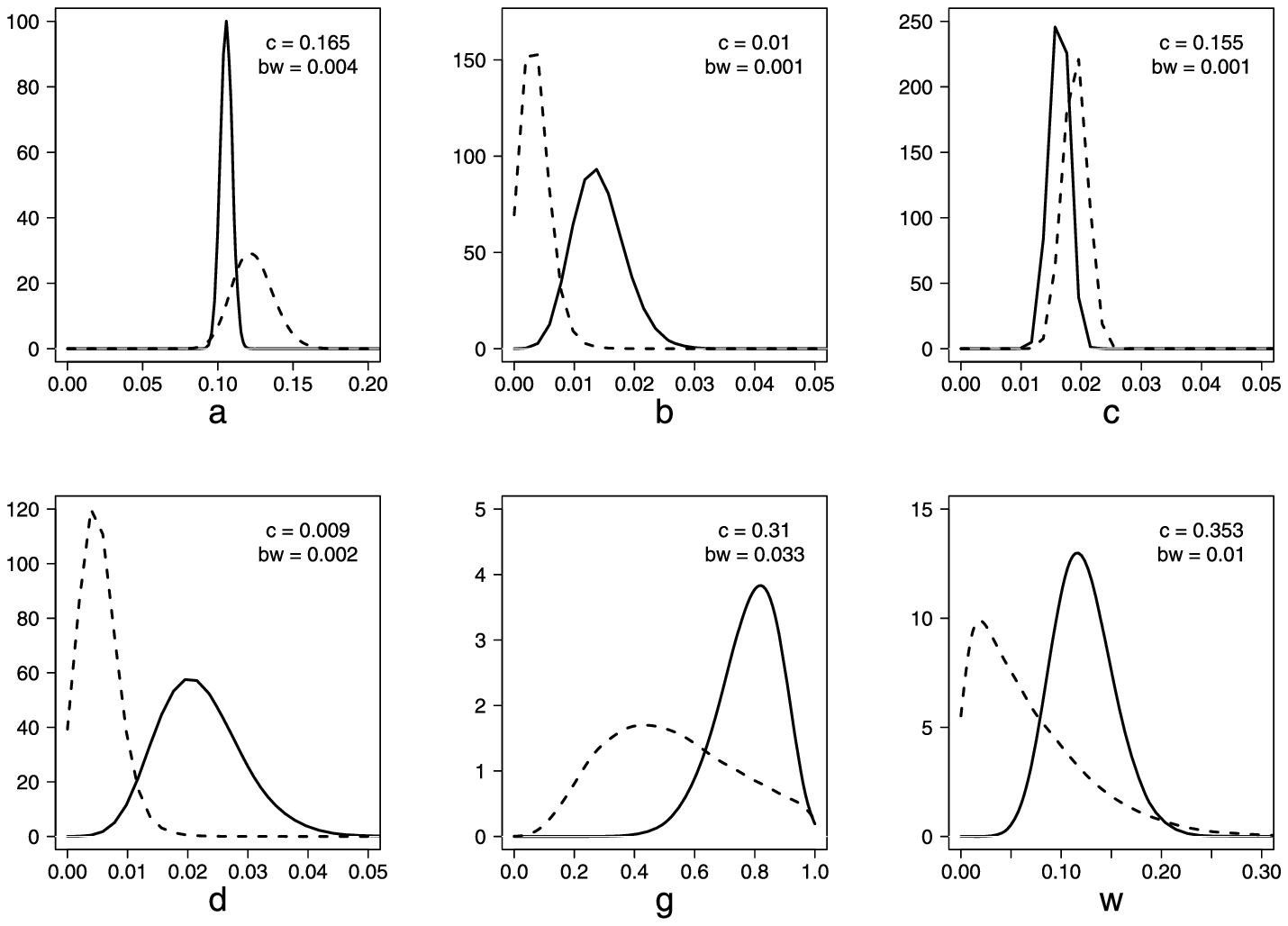}

\caption{The HIV example: posterior
distributions reflecting direct (solid lines) vs indirect (dashed
lines) evidence at the 6 basic parameters with direct evidence
available, $a, \ldots, d, g, w$. The conflict $p$-value ($c$), calculated
as twice the proportion of MCMC samples where $\delta\geq0$ or
$\delta< 0$, whichever is smaller, is given in each plot. The
bandwidth ($bw$) of the kernel density estimate used to plot the
posterior distributions is also shown.}\label{fig_HIVnsAconflicts}
\end{figure*}

\begin{figure*}

\includegraphics{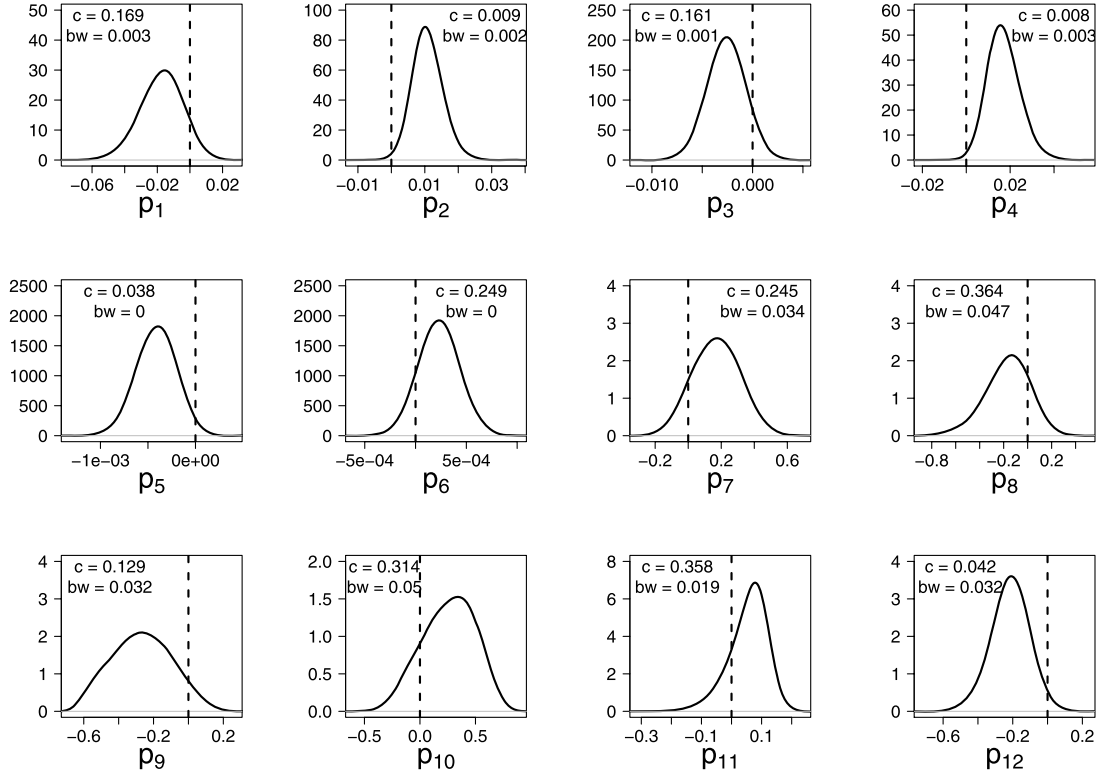}

\caption{The HIV example: the posterior
distribution of the difference function $\delta$ (solid lines) at each
functional parameter $p_i, i \in1, \ldots, 12$ (data-level
cross-validation). $\delta= 0$ is shown by the dashed vertical line.
The conflict $p$-value ($c$), calculated as twice the proportion of MCMC
samples where $\delta\geq0$ or $\delta< 0$, whichever is smaller, is
given in each plot. The bandwidth ($bw$) of the kernel density estimate
used to plot the posterior distributions is also shown.}\label{fig_HIVnsBdelta}
\end{figure*}

Comparison of the direct and indirect evidence informing each $\theta
\in\{a,b,c,d,g,w\}$ is shown in Figure~\ref{fig_HIVnsAconflicts},
together with the conflict $p$-values. The plots and the $p$-values
indicate that the direct evidence informing $b$ and $d$, {that
is,} studies 2 and 4, appear to be in conflict with the rest of the
model. This is confirmed by the second type of node-split, as shown in
Figure~\ref{fig_HIVnsBdelta}: the difference function $\delta_i =
p_{ia} - p_{ib}$ is plotted, and the line $\delta_i = 0$ is shown
together with the conflict $p$-value. Again, studies 2 and 4 are clearly
in conflict with the rest of the model.

%

\subsection{Influenza Example} \label{sec_flu}

\begin{figure*}

\includegraphics{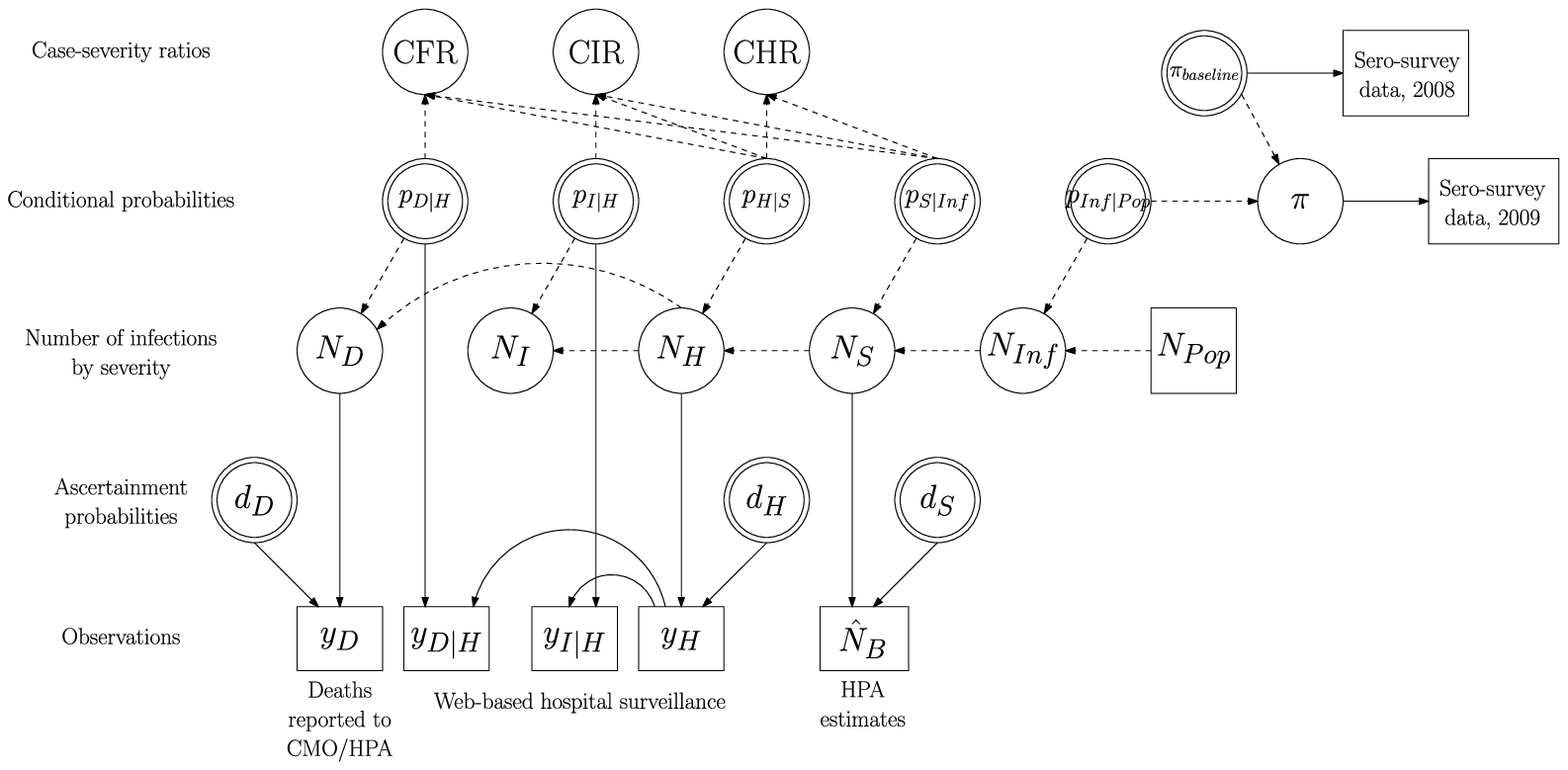}

\caption{The influenza example: DAG for a single
age-group (age index dropped for clarity) of the model reported in
Presanis et~al. (\citeyear{PresanisEtAl2011a}).}\label{fig_fluDAG1}
\end{figure*}

In \citet{PresanisEtAl2011a}, the severity of the first two waves of
the 2009 pandemic influenza outbreak experienced by England was
estimated via a Bayesian evidence synthesis. Severity was measured by
``case-severity ratios'', the probability that an infection will lead to
a more severe event such as hospitalisation (case-hospitalisation ratio
$\mathrm{CHR}$), intensive care (ICU) admission (case-ICU ratio $\mathrm{CIR}$) or death
(case-fatality ratio $\mathrm{CFR}$). Restricting attention to the first wave
only, Figure~\ref{fig_fluDAG1} is a DAG of the model, shown for a
single age group (with the age index dropped for clarity). At the top,
we have the three case-severity ratios, which can each be expressed as
a product of component conditional probabilities $p_{i|j}$ of being a
case at severity level $i$ given severity level~$j$. The
severity\vadjust{\goodbreak}
levels are infection $\operatorname{Inf}$, symptomatic infection $S$, hospitalisation
$H$, ICU admissions $I$ and death $D$, with $\operatorname{Pop}$ denoting the total
(age-group specific) population of England. The case-severity ratios
are functional parameters,\break whereas the probabilities $p_{i|j}$ are
basic parameters assigned prior distributions. At the third layer of
the DAG, the number of infections at each severity level $N_i, i \in\{
\operatorname{Inf},S,H,I,D\}$ is a function of the conditional probabilities $p_{i|j}$
and the number at the less severe level $j$, $N_j$: $N_i = p_{i|j}N_j$.

The authors synthesised data, $y_i$, from surveillance systems
monitoring influenza at different levels of severity, $i$ (Figure~\ref{fig_fluDAG1}), as well as serial sero-prevalence data indirectly
informing the cumulative incidence of infection $p_{\operatorname{Inf}|\operatorname{Pop}}$. Some of
the observed data were recognised to under-ascertain the true number of
infections at different severity levels, so were modelled as
\[
y_i \sim\operatorname{Binomial}(N_i, d_i),
\]
where the probability parameters $d_i$ represent the
under-ascertainment or bias parameters. Estimates $\hat{N}_B$ of the
number symptomatic were produced by the Health Protection Agency (HPA)
from data on primary care consultations for influenza-like-illness and
virological testing for the pandemic strain, adjusted for the
proportion of symptomatic cases contacting primary care. Early on in
the pandemic, these estimates were thought to be underestimates ($B$
here stands for bias). The HPA estimates were therefore incorporated in
the synthesis by modelling them as
\[
\hat{N}_B \sim\operatorname{Binomial}(N_S,
d_S).
\]

Informative Beta priors were given to the probabilities $p_{H|S}$ and
$p_{S|\operatorname{Inf}}$, representing estimates from a cure-rate model fitted to
data on the first few thousand confirmed cases and estimates from the
literature on seasonal influenza, respectively. The remaining
probabilities $p_{i|j}$ were assigned $\operatorname{Beta}(1,1)$ priors. An informative
Beta prior representing estimates from a capture-recapture study was
adopted for the ascertainment probability for death, $d_D$. The other
ascertainment probabilities, $d_H$ and $d_S$, were given $\operatorname{Beta}(1,1)$ priors.

\begin{figure*}

\includegraphics{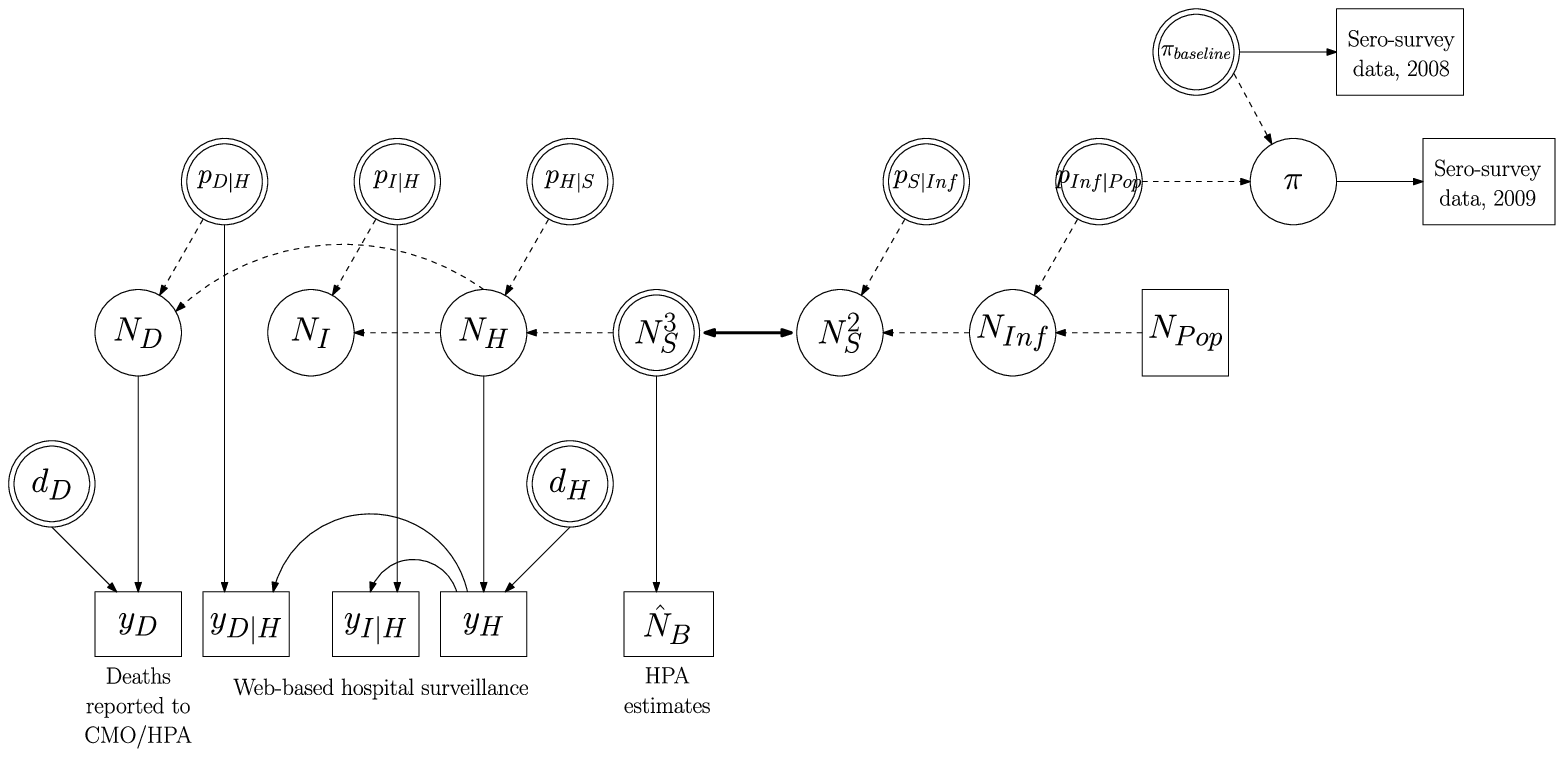}

\caption{The influenza example: the node split at
$N_S$. On the right is model 2, the ``parent'' model. On the left is
model 3, the ``child'' model. The double-headed arrow represents the
comparison between the two.}\label{fig_fluDAG23}
\end{figure*}

Given uncertainties over possible biases in both the sero-prevalence
data and the HPA estimates $\hat{N}_B$, \citet{PresanisEtAl2011a}
carried out a number of sensitivity analyses to the inclusion of these
``denominator'' data, making different assumptions about which source, if
any, is biased. An alternative approach is to split\vadjust{\goodbreak} the DAG at the node
$N_S$, having dropped from the model for now the bias parameter $d_S$,
to assess the consistency of the different groups of evidence. Denote
the full model of Figure~\ref{fig_fluDAG1} but with $d_S$ removed by
Model 1. We wish to compare the sero-prevalence data combined with the
informative prior for the proportion symptomatic $p_{S|\operatorname{Inf}}$ (the
``parent'' model 2) against the HPA estimates $\hat{N}_B$, combined with
all the severe end data and priors (the ``child'' model 3). Figure~\ref{fig_fluDAG23} shows this node-split.
In the child model, $\log(N_S^3)$ is assigned the Jeffreys' prior
(uniform on the real line). The difference function we are interested
in is $\delta= \log(N_S^2) - \log(N_S^3)$ and the tail probability we
assess is the one-sided probability $c = \operatorname{Pr}\{\delta< 0\}$, since the
HPA estimates are recognised underestimates. The $p$-values are
calculated as the proportion of MCMC samples where $\delta< 0$.

%

\begin{figure*}

\includegraphics{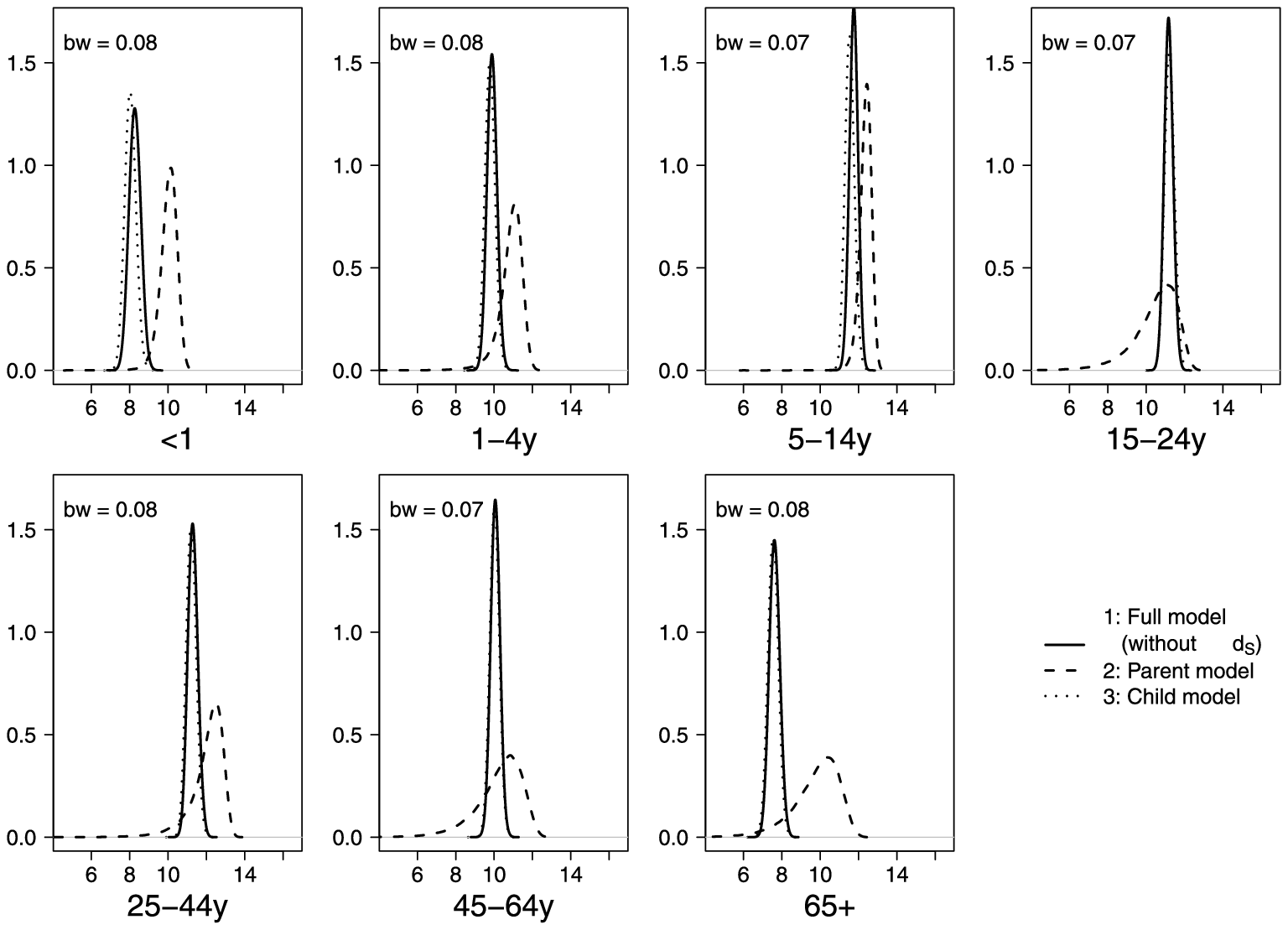}

\caption{The influenza example: the posterior
distribution of the number symptomatic $N_S$ on the log-scale, by age
group and model. The bandwidth ($bw$) of the kernel density estimate
used to plot the posterior distributions is shown in the top left of
each plot.}\label{fig_fluLnN}
\end{figure*}

\begin{figure*}

\includegraphics{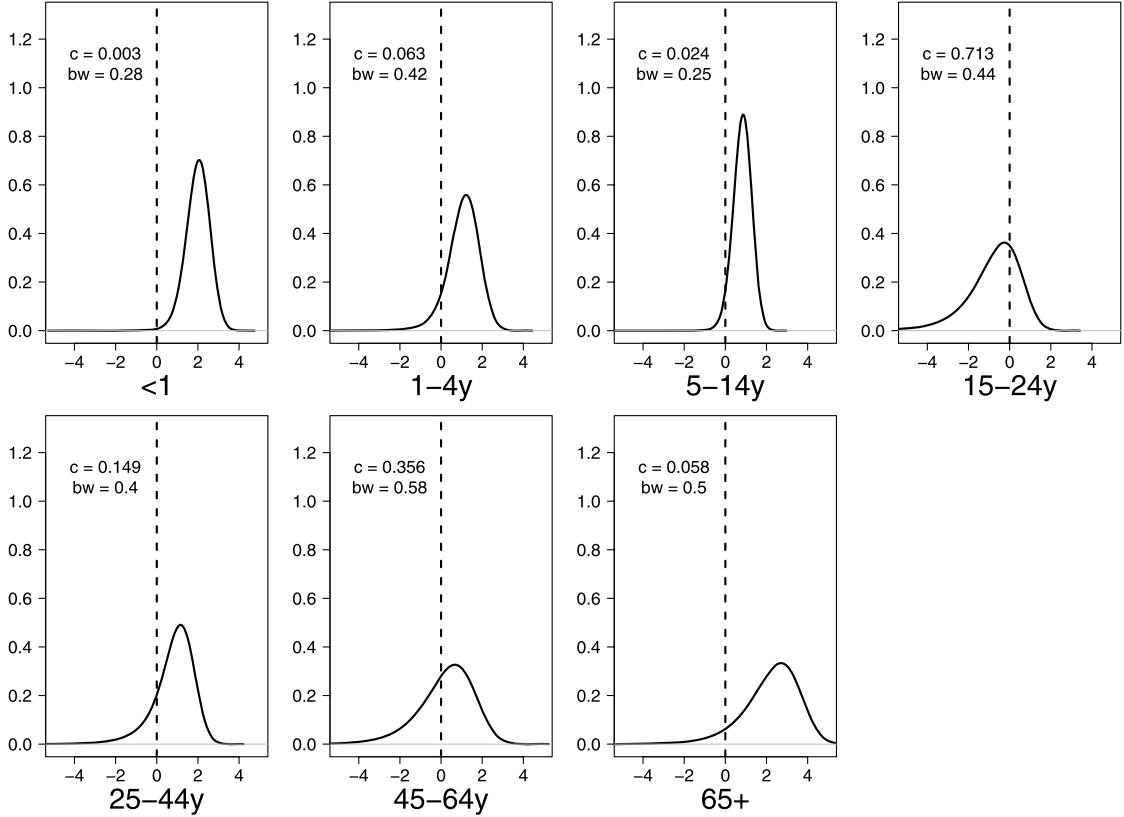}

\caption{The influenza example: the posterior
distribution of the difference function $\delta= \log(N_S^2) - \log
(N_S^3)$. The vertical dashed line gives $\delta= 0$. The one-sided
conflict $p$-value ($c$), calculated as the proportion of MCMC samples
where $\delta< 0$, is given in each plot. The bandwidth ($bw$) of the
kernel density estimate used to plot the posterior distributions is
also shown.}\label{fig_fluDelta}
\end{figure*}

Three chains of a million MCMC iterations each, with the first 200,000
discarded as burn-in, were thinned to every $10$th iteration
to obtain 240,000 posterior samples on which to base results. Figure~\ref{fig_fluLnN} shows, for each of seven age groups, the posterior
distributions of $\log(N_S)$ in each of the three models: the full
model~1, the parent model 2 and the child model 3.
Two facts are immediately apparent from these plots: first, that the
estimates provided by the sero-prevalence data in the parent model are
very uncertain, particularly in the adult age groups, where sample
sizes were small; and second, the resulting influence of the more
severe end data, including the HPA estimates, is seen in the closeness
of the full model estimates to the child model estimates. What is less
apparent from these plots is the extent of conflict between the two
sets of information, except in the age group $<1$ where the conflict is
clear to see. The plots of the difference function $\delta$ in Figure~\ref{fig_fluDelta},
together with the conflict $p$-values shown, are
required to assess the conflict in other age groups.
From these, we see $p$-values less than $0.1$ in the child age groups,
providing evidence of conflict in these groups. By contrast, in the
adult age groups, the uncertainty in the estimates of $N_S^2$ in the
parent model~2 is so large that there is no conflict.\vadjust{\goodbreak}

%

\begin{figure*}

\includegraphics{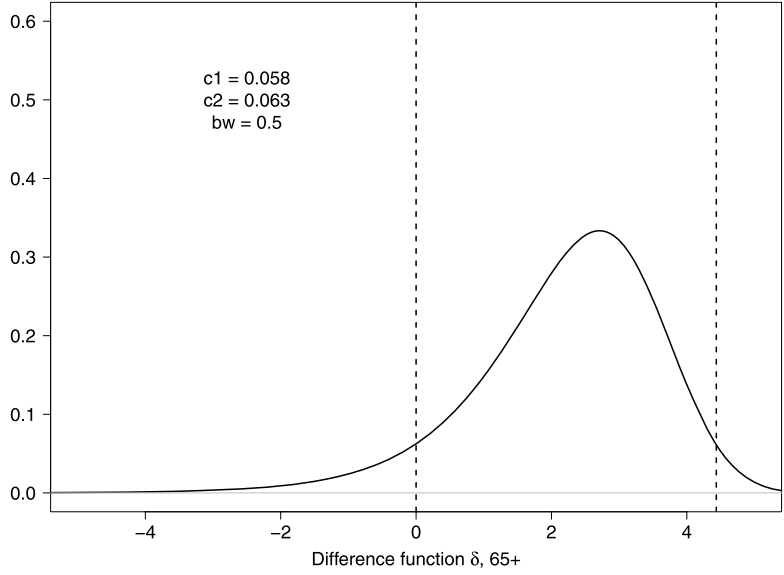}

\caption{The influenza example: one- ($c_1$) and
two-sided ($c_2$) $p$-values for the $65+$ age group, calculated using
kernel density estimation with bandwidth $bw$. The two vertical dashed
lines show where $\delta= 0$ and the corresponding value $\delta= k =
4.44$, such that the density at 0 and at $k$ is equal, lie in the
posterior distribution of $\delta$.}\label{fig_flu2sided}
\end{figure*}

If we had not already suspected the HPA estimates were underestimates
and wished to assess potential conflict using a two-sided test, then in
this example calculation of the two-sided $p$-value would not be so
straightforward. Particularly in the over-65 age group, the posterior
difference function is skewed (Figure~\ref{fig_fluDelta}). We could
therefore, as suggested in Section~\ref{sec_compare}, use kernel
density estimation to calculate the two-sided $p$-value. Figure~\ref{fig_flu2sided} compares the one-sided to the resulting two-sided
$p$-value for the $65+$ group, using a bandwidth of 0.5.
%

\subsection{Multivariate Example: Growth in Rats} \label{sec_rats}
\begin{figure}

\includegraphics{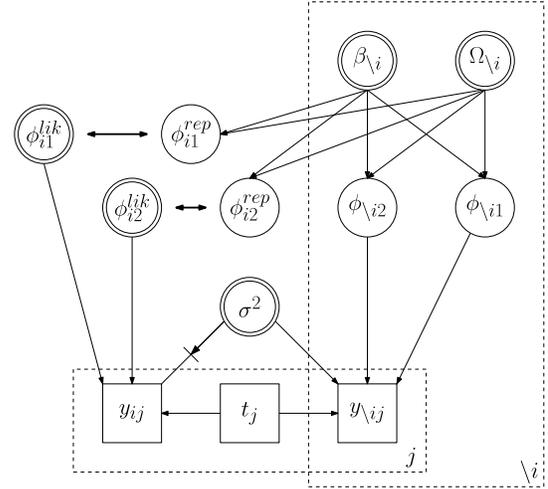}

\caption{The rats example: DAG showing the
comparison of the fixed effect model for rat $i$ ($\bolds{\phi}
_i^{\mathrm{lik}}$) with the random effects prediction from the remaining rats
($\bolds{\phi}_i^{\mathrm{rep}}$).}\label{fig_ratsDAG}
\end{figure}

A simpler example that nevertheless demonstrates the complexity of
multivariate node-splitting is provided by data from a population
growth experiment considered by \citet{GelfandEtAl1990}. The data
comprise the weight $y_{ij}$ of each of 30 rats ($i \in1, \ldots, 30$)
at ages 8, 15, 22, 29 and 36 days, indexed by $j \in1, \ldots, 5$. The
authors' null model $H_0$ assumes a normal error and random coefficient
linear growth curves with time $t_j$ measured in days centred on 22.
The intercept and gradient are given a bivariate normal prior, so that
%
%
\begin{eqnarray}\label{eq_rats}
y_{ij} & \sim& \mathrm{N}\bigl(\mu_{ij},
\sigma^2\bigr),
\nonumber
\\
\mu_{ij} & = & \phi_{i1} + \phi_{i2}
t_{j},
\\
\bolds{\phi}_i & \sim& \operatorname{MVN}_2(\bolds{\beta},
\bolds{\Omega}),
\nonumber
\end{eqnarray}
where $\bolds{\phi}_i = (\phi_{i1}, \phi_{i2})^T$ and $\bolds
{\beta}, \bolds{\Omega}, \sigma^2$ are given proper but very
diffuse prior distributions:
\begin{eqnarray*}
\bolds{\beta} & \sim& \operatorname{MVN}_2 \left( \pmatrix{0
\cr
0 }
, \pmatrix{
10^{-6} & 0
\cr
0 & 10^{-6} } \right),
\\
\boldsymboll{W} & \sim& \operatorname{Wishart}\left(\pmatrix{ 200 & 0
\cr
0 & 0.2 }, 2 \right),
\\
\bolds{\Omega} & = & \boldsymboll{W}^{-1},
\\
\tau& \sim& \Gamma\bigl(10^{-3},10^{-3}\bigr),
\\
\sigma^2 & = & \tau^{-1}.
\end{eqnarray*}
We wish to examine the suitability of the bivariate random effects
distribution for each rat, which in this case requires assessing a
multivariate node-split, at the vector of parameters $\bolds{\phi}
_{i}$ (Figure~\ref{fig_ratsDAG}). For each rat $i$ (i.e., each
cross-validation), in one partition, we fit a fixed effects model to
the data $\boldsymboll{y}_{i}$ to estimate the nodes denoted $\bolds
{\phi}_{i}^{\mathrm{lik}}$, which are assigned independent (improper) Jeffreys'
priors. In the other partition, we predict the nodes $\bolds{\phi}
_{i}^{\mathrm{rep}}$ from the random effects model fitted to the data on the
remaining rats, denoted by $\boldsymboll{y}_{\setminus{i}}$.
As with the example from \citet{MarshallSpiegelhalter2007}, to form a
complete split in the DAG would require also splitting the variance
parameter $\sigma^2$. Since our primary interest is in assessing the
random effects distribution for $\bolds{\phi}_i$ rather than for
the rat-specific data $\boldsymboll{y}_{i}$, and as a rat-specific
variance $\sigma^2_i$ may not be well identified from the data on one
rat alone, we treat $\sigma^2$ as a nuisance parameter. We therefore
place a ``cut'' in the DAG to prevent feedback from rat $i$'s data to
$\sigma^2$.

\begin{figure*}

\includegraphics{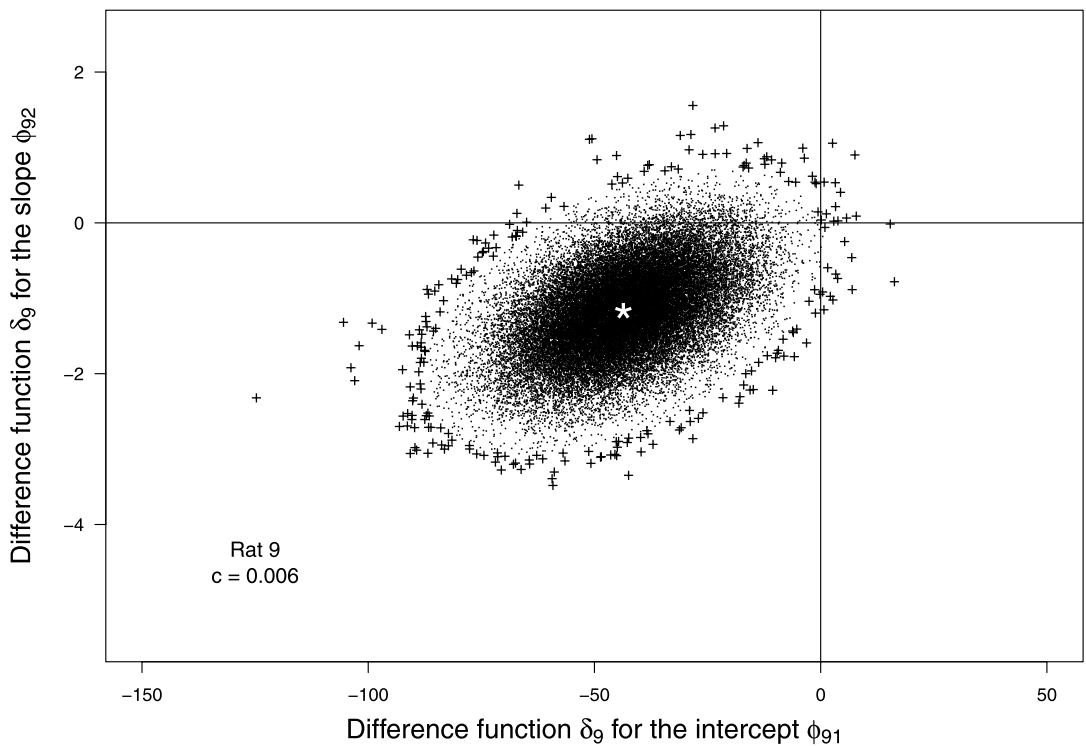}

\caption{The rats example: joint posterior
distribution of $\bolds{\delta}_{9}$. Points more extreme than
$(0,0)$ (i.e., further from the mean in terms of Mahalanobis distance
and therefore lying in the tail of the distribution) are shown as
crosses. The white star denotes the posterior mean $\mathbb
{E}_p(\bolds{\delta}_9)$.}\label{fig_rat9}
\end{figure*}

\begin{figure*}

\includegraphics{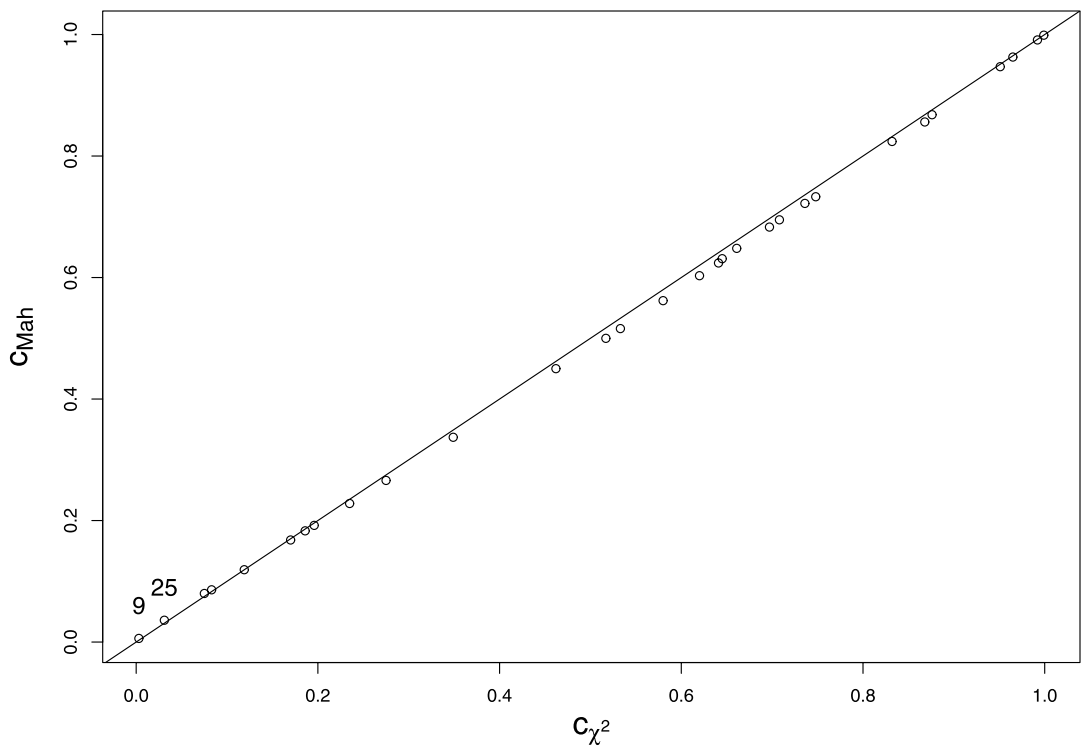}

\caption{The rats example: on the x-axis,
$c_{\mathcal{\chi}^2}$ is the $p$-value using the $\mathcal{\chi}^2$
approach; on the y-axis, $c_{Mah}$ is the $p$-value using the Mahalanobis
approach. Rats 9 and 25 have $p$-values less than 0.05.}\label{fig_ratPvals}
\end{figure*}

A multivariate difference function was defined for each rat $i$:
$\bolds{\delta} _i= \bolds{\phi}_i^{\mathrm{rep}} - \bolds{\phi}
_i^{\mathrm{lik}}$ and MCMC samples from the posterior distribution of
$\bolds{\delta}_i$ were obtained based on two chains of $20\mbox{,}000$
iterations each, following a burn-in of $20\mbox{,}000$ iterations. Plots for
each rat of the samples from the joint posterior distribution of
$\bolds{\delta}_{i}$ suggest that at least uni-modality and
symmetry hold approximately, and possibly bivariate normality also
(see, e.g., rat 9 in Figure~\ref{fig_rat9}). We therefore
calculate a conflict $p$-value based on each of the first two suggestions
in Section~\ref{sec_multi}, shown in Figure~\ref{fig_ratPvals}. Both
methods of defining the $p$-value give similar results and suggest that
rat 9 is discrepant, with $p$-values of 0.003 and 0.006 for the $\mathcal
{\chi}^2$- and Mahalanobis-based methods, respectively. Figure~\ref{fig_rat9} shows the joint posterior distribution of $\bolds{\delta}
_{9}$, with crosses denoting samples that are further away (in terms
of Mahalanobis distance) from the posterior mean (the white star) than
is the point $(0,0)$.\looseness=1
%

A parallel literature exists on model diagnostics to identify
unit-level outliers for classical multilevel models \citep
{LangfordLewis1998}. The basic idea of this diagnostic is to add extra
fixed effects or dummy variables for the unit under consideration and
compare this model and the null by comparing the change in deviance
between the two, evaluated at the maximum likelihood estimate, to a
$\mathcal{\chi}^2_1$ distribution. For example, in the rats study, an
extra fixed effect could be added for the slope and intercept of rat
$i$ and this model (the alternative) could be compared with the null
model~(\ref{eq_rats}). \citet{OhlssenEtAl2007} show that this method is
equivalent to the Bayesian cross-validatory mixed-predictive $p$-value of
\citeauthor{MarshallSpiegelhalter2007}\break (\citeyear{MarshallSpiegelhalter2007}) in balanced one-way random effects
models with uniform priors.

\section{Discussion} \label{sec_discuss}

We have described a generic simulation-based technique that can be used
as a diagnostic for conflicting information at any node(s) $\bolds
{\theta}$ in a DAG, generalising the conflict $p$-value first proposed by
\citet{MarshallSpiegelhalter2007}. We have presented a framework
focussed on a conflict diagnostic that is both useful and
straightforward to construct for complex, typically nonstandard,
evidence synthesis models. We have given recommendations, via three
substantive examples, for how to perform node-splits to assess
different types of conflict. In particular, we have demonstrated
different methods of handling multivariate node-splits, dependent on
the context. If a ``separator'' parameter is not of primary interest, but
a nuisance parameter, we suggest making use of the ``cut'' function in
software such as OpenBUGS, to prevent information flow to the nuisance
parameter from one partition. For multivariate ``separator'' nodes that
are of importance, we recommend a hierarchy of options for defining a
multivariate conflict $p$-value, dependent on the normality, uni-modality
and symmetry or otherwise of the posterior distribution of the
difference function. In focussing on nonstandard but realistic
situations, our framework goes beyond that so far proposed in the
literature [e.g., \cite{BayarriCastellanos2007}; \cite{GasemyrNatvig2009}].

There are still practical and theoretical considerations raised by the
analysis of conflict that require further investigation. For large
complex models, a systematic examination of every potential conflict at
every node in a DAG may induce a large computational cost, since---as\vadjust{\goodbreak}
with cross-validation---every node-split requires a new model run.
Furthermore, such a systematic conflict assessment would result in the
multiple comparison problem. To address these issues, different
approaches may be taken. In practice, approximations to full
cross-validation as suggested by \citet{MarshallSpiegelhalter2007} may
be employed. Or it may be prudent to be selective about node-splits to
examine, based on the context of the problem, as we were in the
influenza example and to some extent in the HIV example. One strategy
may be to employ a diagnostic such as comparison of posterior mean
deviance to the number of data items (\cite
{Dempster1997}; \cite{SpiegelhalterEtAl2002}) to detect lack of fit to
particular items, which may then be an indicator of conflicting
evidence (\cite{AdesCliffe2002}; \cite{PresanisEtAl2008}). Then the choice of
node-splits to examine may be guided by the locations in the DAG of
lack of fit. However, the posterior mean deviance has its own
limitations as a diagnostic \citep{PresanisEtAl2008} and, indeed,
conflicting evidence may not necessarily manifest as lack of fit. This
was the case with the influenza example (results not shown), where
instead an informative prior for the proportion of infections which
were symptomatic was shifted in the posterior to an implausibly low
range. If systematic conflict assessment is indeed an aim of the
analyst, then the problem of multiple testing may be addressed by
combining node-splitting with methods for controlling the false
discovery rate [FDR, \cite{BenjaminiHochberg1995}; \cite{JonesEtAl2008}]. A
further complication is the possibility of correlated hypothesis tests,
since the different node-splits assessed may have nodes in common with
each other. FDR methods do extend to correlated tests \citep
{BenjaminiYekutieli2001}, but such methods may have an impact on the
power to detect conflict or identify the network of nodes in a DAG that
are conflicting (\cite{HothornEtAl2008}; \cite{BretzEtAl2011}). A possibility
for further investigation is to account for correlation through the
general framework for multiple comparisons of \citet{HothornEtAl2008}, \citet{BretzEtAl2011}, although this requires asymptotic
multivariate normality.

In models where flat improper priors are employed and the likelihood
dominates the prior, the posterior will be approximately normal, so
that the conditions of the framework of \citet{GasemyrNatvig2009} hold,
and the conflict $p$-values will therefore be uniformly distributed under\vadjust{\goodbreak}
the null hypothesis that $\theta_a = \theta_b$. However, in practice,
analysts modelling complex phenomena rarely use improper priors,
perhaps instead using either (i) proper priors with very large
variances; or (ii) informative priors to represent previous knowledge
or to ensure identifiability. In case (i), approximate normality of the
posterior difference function $\delta$ will again ensure the $p$-values
are uniform under the null, though sensitivity analyses should be
employed to check the prior is dominated by the likelihood.
Furthermore, our recommendation of the use of Jeffreys' priors for
appropriate transformations of $\theta_b$ should, we hope, result in a
posterior difference function that is at least approximately
(multivariate) normal. Sensitivity analysis may again be employed to
assess our choice of reference prior (Jeffreys' or some other uniform
prior) and transformation function $h(\cdot)$ for the split node. In
case (ii), where informative priors are assigned to nodes other than
$\theta_b$, the posterior distribution of $\theta_a$ and hence the
difference function $\delta$ are not guaranteed to be even
approximately normal. This potential nonnormality may be exacerbated
by small sample sizes, as was the case in our influenza example, where
in some age groups the posterior difference function was somewhat
skewed. Our suggestion is to then use kernel density estimation to
obtain a conflict $p$-value, though clearly more work would be required
to understand the distribution of such a $p$-value. Kernel density
estimation also raises its own questions of how to choose a kernel and
a bandwidth with which to smooth, suggesting the need for sensitivity
analyses to these choices. Kernel density estimation is also
computationally demanding, particularly for multivariate surfaces, and
may result in $p$-values that are not invariant to transformation. More
generally, a conflict measure defined as the probability that a density
function is smaller than a given threshold is preferable in cases where
the distribution of $\delta$ is not symmetric and unimodal, since it
would identify regions of surprise not only in the tails, but also in
shallow anti-modes. However, the question of how to obtain such a
$p$-value, that is also invariant to transformation, requires further
investigation \citep{EvansJang2010}.

Our case studies in assessing conflict have demonstrated the great
importance of visualisation, with plots of, for example, the posterior
difference function or the posterior distributions corresponding to
different partitions of a DAG adding to our understanding of what and
where conflict is occurring. The influence of different partitions of
evidence on estimation can also be visualised from such plots. It is
important to note that node-splitting is a diagnostic, one step in the
inference-criticism cycle and a pointer to where in a DAG further
analysis is required. The next step is to understand, within the
context of the specific problems under analysis, the reasons for the
inconsistencies, and therefore to resolve the conflict. There are many
possibilities for accommodating conflict, including: the exclusion of
sources of evidence; the addition of extra variation to account for
potential bias (e.g., \cite{AndradeOHagan2006}; \cite{EvansJang2011});
and model elaboration to explicitly model biases or account for
unexplained heterogeneity (e.g.,
\cite{DuMouchelHarris1983}; \cite{LuAdes2006};
\cite{Greenland2009};
\cite{PresanisEtAl2008};
\cite{WeltonEtAl2009}; \cite{TurnerEtAl2009};
\cite{HigginsEtAl2012};
\cite{WhiteEtAl2012}).
Any such model development will then lead to the next iteration of
inference and criticism, in the spirit of \citet{Box1980} and \citet
{OHagan2003}. Clearly, also, in any Bayesian analysis, sensitivity
analyses, both to prior distributions, whether vague or informative,
and in the case of evidence synthesis, to the sources of information
included, are an important part of the model criticism process.

\section*{Acknowledgements}

This work was supported by the Medical Research Council [Unit Programme
Numbers U105260566 and U105260557]. We thank the Health Protection
Agency for providing the data for the influenza example and Richard
Pebody, in particular, for useful discussions about the detected
conflict. We are also grateful to the participants of the workshop
``Explaining the results of a complex probabilistic modelling exercise:
conflict, consistency and sensitivity analysis'', sponsored by the
Medical Research Council Population Health Sciences Research Network,
in Cambridge in 2006, for many useful discussions. We acknowledge also
the constructive comments of two referees, that led to an improved manuscript.


%



\end{document}